\documentclass[usenatbib]{mn2e}
\usepackage{times,amssymb,amsbsy}
\usepackage{graphicx}

\newcommand{\D}[2]{\frac{\partial #2}{\partial #1}}
\newcommand{\DD}[2]{\frac{\partial^2 #2}{\partial #1^2}}
\newcommand{\deriv}[2]{\frac{{\rm d} #2}{{\rm d} #1}}
\newcommand{\dderiv}[2]{\frac{{\rm d}^2 #2}{{\rm d} #1^2}}
\newcommand\bb[1]{\mbox{\boldmath{$#1$}}}
\newcommand\grad{\bb{\nabla}}
\newcommand\bcdot{\bb{\cdot}}
\newcommand\btimes{\bb{\times}}
\newcommand{\mc}[1]{\mathcal{#1}}

\newcommand{\msb}[1]{\bb{\mathsf{#1}}}

\newcommand{\imag}{{\rm i}}
\newcommand{\eb}{\hat{\bb{e}}_b}
\newcommand{\ev}{\hat{\bb{e}}_v}
\newcommand{\ex}{\hat{\bb{e}}_x}
\newcommand{\ey}{\hat{\bb{e}}_y}
\newcommand{\ez}{\hat{\bb{e}}_z}
\newcommand{\vasq}{v^2_{\rm A}}

\newcommand{\lhmi}{\Lambda_\mathrm{H}^{-1}}
\newcommand{\lh}{\ell_{\rm{H}}}

\title[Hall-dominated MRI turbulence]
{Magnetic self-organisation in Hall-dominated magnetorotational turbulence}
\author[M.~W. Kunz \& G.~Lesur]
{Matthew W. Kunz$^{1}$\thanks{NASA Einstein Postdoctoral Fellow} and Geoffroy Lesur$^{2}$\thanks{E-mail: geoffroy.lesur@ujf-grenoble.fr} \\
$^{1}$ Department of Astrophysical Sciences, 4 Ivy Lane, Peyton Hall, Princeton University, Princeton, NJ 08544, U.~S.~A. \\
$^{2}$ UJF-Grenoble 1 / CNRS-INSU, Institut de Plan\'{e}tologie et d'Astrophysique de Grenoble (IPAG) UMR 5274, Grenoble, F-38041, France
}

\date{}
\pagerange{\pageref{firstpage}--\pageref{lastpage}} \pubyear{2013}
\def\LaTeX{L\kern-.36em\raise.3ex\hbox{a}\kern-.15em
    T\kern-.1667em\lower.7ex\hbox{E}\kern-.125emX}
\begin{document}
\label{firstpage} \maketitle

\begin{abstract}
The magnetorotational instability (MRI) is the most promising mechanism by which angular momentum is efficiently transported outwards in astrophysical discs. However, its application to protoplanetary discs remains problematic. These discs are so poorly ionised that they may not support magnetorotational turbulence in regions referred to as `dead zones'. It has recently been suggested that the Hall effect, a non-ideal magnetohydrodynamic (MHD) effect, could revive these dead zones by enhancing the magnetically active column density by an order of magnitude or more. We investigate this idea by performing local, three-dimensional, resistive Hall-MHD simulations of the MRI in situations where the Hall effect dominates over Ohmic dissipation. As expected from linear stability analysis, we find an exponentially growing instability in regimes otherwise linearly stable in resistive MHD. However, instead of vigorous and sustained magnetorotational turbulence, we find that the MRI saturates by producing large-scale, long-lived, axisymmetric structures in the magnetic and velocity fields. We refer to these structures as {\it zonal fields} and {\it zonal flows}, respectively. Their emergence causes a steep reduction in turbulent transport by at least two orders of magnitude from extrapolations based upon resistive MHD, a result that calls into question contemporary models of layered accretion. We construct a rigorous mean-field theory to explain this new behaviour and to predict when it should occur. Implications for protoplanetary disc structure and evolution, as well as for theories of planet formation, are briefly discussed.
\end{abstract}


%
%
\section{Introduction}\label{sec:intro}

Protoplanetary discs are poorly ionised. This fact casts doubt upon whether the most promising mechanism for enhanced angular-momentum transport in accretion discs, the magnetorotational instability (MRI; \citealt{bh91,bh98}), is capable of driving the observationally inferred mass-accretion rates in these systems \citep[e.g.][]{hcgd98}. Not only are the most potent sources of ionisation (e.g.~cosmic rays, stellar X-rays, UV radiation) shielded over significant portions of these discs, but also the presence of dust grains is anticipated to remove an appreciable fraction of the charge from the gas phase. All this makes it improbable that protoplanetary discs are magnetically coupled across their full radial and vertical extents (for a review, see \citealt{armitage11}). This is particularly true in the dense midplane, where the principal source of ionisation is likely to be weak radioactivity.

There are reasons to believe, however, that such pessimism is unwarranted. The MRI appears to be much more resilient in the face of diffusive losses than one may at first suspect. While a low degree of ionisation  is known to decouple the MRI-unstable charged species from the bulk neutral fluid in processes known as ambipolar diffusion \citep{bb94,hs98,kb04,desch04,bs11,sbsab13} and Ohmic dissipation \citep{jin96,sm99,fsh00}, it is somewhat surprising that the critical ionisation fraction at $\sim$$1~{\rm au}$ is just $\sim$$10^{-13}$ for typical protoplanetary discs \citep[e.g.][]{balbus11}. This is because the charged species have $\sim$$1~{\rm yr}$ to communicate the magnetic field to the neutrals via collisions. This renders the innermost ($r \lesssim 0.1~{\rm au}$), outermost ($r \gtrsim 30~{\rm au}$), and surface-layer ($z \gtrsim 0.1~{\rm au}$) regions magnetically active, either by thermal ionisation of metals or by unshielded ionising radiation.

Of course, these numbers come with large uncertainties, and much of the research concerning the MRI in protoplanetary discs has boiled down to determining the extent of magnetically active regions by coupling chemical networks of increasing complexity to either linear stability analyses \citep[e.g.][]{gammie96,ig99,smun00,sw03,sw05,sw08,ws12} or some sort of nonlinear criterion based upon numerical experiments \citep[e.g.][]{ftb02,in06,bg09,bai11}. However, while determining the chemical abundances and consequent diffusivities in such discs is without a doubt essential to improving our understanding of disc stability, structure, and evolution, the results presented in this Paper suggest that the philosophy driving this approach can be misleading. 

Here, we take an alternative route to understanding magnetorotational turbulence in protoplanetary discs. We forego a detailed study of disc chemistry and instead concentrate on the turbulent disc dynamics themselves. Employing nonlinear numerical simulations and mean-field theory, we investigate the impact of Ohmic dissipation and the Hall effect on magnetorotational turbulence. We extend previous work by Sano \& Stone (2002a,b\nocite{ss02a,ss02b}; hereafter, SS02) into the Hall-dominated regime, and obtain qualitatively new results. Instead of vigorous and sustained magnetorotational turbulence, we find that the Hall-MRI saturates by producing large-scale, long-lived, axisymmetric (`zonal') structures in the magnetic and velocity fields. Their emergence---a result of the anti-diffusive nature of the Hall effect when the Maxwell stress increases with magnetic-field strength---causes a reduction in turbulent transport by at least two orders of magnitude from extrapolations based upon resistive MHD. 

Our results suggest that existing estimates of the depth of magnetorotationally active layers in protoplanetary discs based on damping by Ohmic dissipation and ambipolar diffusion are likely to be in error. This conclusion has been reached before by other authors \citep{ws12}, but for different reasons. Those authors put emphasis on the fact that the Hall effect can render a disc linearly unstable \citep{wardle99,bt01} even in the presence of strong Ohmic and ambipolar diffusion. In this case, the critical magnetic Reynolds number ${\rm Rm}_{\rm crit}$ for magnetorotational turbulence ought to be smaller than is often assumed. By contrast, our results suggest that, even when a disc is deemed magnetically active from the perspective of linear analysis, the actual turbulent transport that results may be much too small to be considered `active'. Even for discs in which ${\rm Rm} \gtrsim 10^3$, an order-of-magnitude larger than what is usually considered the critical value, the Hall effect can cause a turbulent bifurcation to a low-transport state. As a result, Hall-dominated regions of protoplanetary discs ($r \sim 5$--$10~{\rm au}$), while magnetically active, may nevertheless exhibit prohibitively low accretion rates.

The paper is organised as follows. In Section \ref{sec:eqns} we present the governing shearing-sheet equations of resistive Hall-MHD. In Sections \ref{sec:hall} and \ref{sec:hallmax} we highlight two key physical concepts encapsulated by these equations---the conservation of canonical vorticity, and the close connexion between the transport of magnetic flux and the transport of angular momentum. Section \ref{sec:ppds} places these considerations in the context of protoplanetary discs, from which we obtain numerical estimates of the four dimensionless free parameters in our system (\S\,\ref{sec:params}). We close Section \ref{sec:prelim} by proving that MRI `channel' modes remain exact nonlinear solutions despite the complicating features of the Hall effect (\S\,\ref{sec:channel}); we defer to Appendix \ref{app:parasite} an investigation of their stability to secondary `parasitic' modes. Section \ref{sec:simulations} presents the numerical approach we have adopted, the tests we have employed to verify its stability and accuracy (see also Appendix \ref{app:hallstab}), and the results of using this approach to study the Hall-dominated MRI in the linear and nonlinear regimes. These results motivate the construction of a mean-field theory that explains both the emergence of zonal structures and the transition to a low-transport state observed in our simulations (\S\,\ref{sec:bifurcation}). Finally, in Section \ref{sec:discussion} we summarise our results and briefly comment on their implications for protoplanetary discs and planetesimal formation.

%
%
\section{Preliminaries}\label{sec:prelim}

%
%
\subsection{Shearing-sheet equations}\label{sec:eqns}

We adopt the shearing-sheet approximation \citep{glb65}, a useful framework for describing phenomena that vary on lengthscales much less than the large-scale properties of the disc. A small patch of the disc, co-orbiting with a fiducial point $r_0$ in the midplane of the unperturbed disc at an angular velocity $\bb{\Omega} = \Omega_0 \ez$, is represented in Cartesian coordinates with the $x$ and $y$ directions corresponding to the radial and azimuthal directions, respectively. Differential rotation is accounted for by including the Coriolis force and by imposing a background linear shear, $\bb{v}_0 = 2 A_0 x \ey$, where
\[
A_0 = \frac{r_0}{2}\left. \deriv{r}{\Omega(r)} \right|_{r=r_0}
\]
is the Oort `A' value; Keplerian rotation yields $A_0 = -(3/4) \Omega_0$. We take the flow to be incompressible, a good assumption when the magnetic pressure is much less than the gas pressure.

The equations of motion are then
\begin{eqnarray}\label{eqn:force}
\lefteqn{
\D{t}{\bb{v}} = - \bb{v} \bcdot \grad \bb{v} - \frac{1}{\rho} \grad P + \frac{\bb{J}\btimes\bb{B}}{c\rho} - 2 \bb{\Omega}_0 \btimes \bb{v} - 4 A_0 \Omega_0 x \ex 
}\nonumber\\*&&\mbox{} 
+ \nu \nabla^2 \bb{v} ,
\end{eqnarray}
\begin{equation}\label{eqn:induction}
\D{t}{\bb{B}} = \grad \btimes \left( \bb{v} \btimes \bb{B} - \frac{ \bb{J} \btimes \bb{B} }{e n_{\rm e}} \right) + \eta \nabla^2 \bb{B} ,
\end{equation}
subject to the constraints
\begin{equation}\label{eqn:divv}
\grad \bcdot \bb{v} = 0 ,
\end{equation}
\begin{equation}\label{eqn:divb}
\grad \bcdot \bb{B} = 0 .
\end{equation}
Our notation is standard: $\rho$ is the (homogeneous) mass density, $\bb{v}$ is the velocity, $P$ is the gas pressure, $\bb{B}$ is the magnetic field, and
\[
\bb{J} = \frac{c}{4\pi} \grad \btimes \bb{B}
\]
is the current density. The number density of electrons $n_{\rm e}$ is taken to be constant and uniform, as are the viscosity $\nu$ and resistivity $\eta$. For future reference, we also introduce the total number density $n = \rho / m$, where $m$ is the mean mass per particle, and the ion mass density $\rho_{\rm i} = m_{\rm i} n_{\rm i}$. Quasi-neutrality (i.e.~$n_{\rm e} = Z n_{\rm i}$) is assumed.

Henceforth, the subscript `0' on $A_0$ and $\Omega_0$ is dropped.

%
%
\subsection{Lorentz force, Hall effect, and canonical vorticity}\label{sec:hall}

In the incompressible approximation, the gas pressure $P$ is determined not by an equation of state, but rather by satisfying the incompressibility condition (\ref{eqn:divv}). In fact, it is customary to eliminate the pressure by taking the curl of equation (\ref{eqn:force}) to obtain an evolutionary equation for the flow vorticity $\bb{\omega} \equiv \grad \btimes \bb{v} + 2 \bb{\Omega}$:
\begin{equation}\label{eqn:vorticity}
\D{t}{\bb{\omega}} = \grad \btimes \left( \bb{v} \btimes \bb{\omega} + \frac{\bb{J} \btimes \bb{B}}{c\rho} \right) + \nu \nabla^2 \bb{\omega} .
\end{equation}
The form of equation (\ref{eqn:vorticity}) is very similar to that of equation (\ref{eqn:induction}). Just as the Lorentz force changes the number of vortex lines threading a fluid element, the Hall effect (represented by the penultimate term in eq.~\ref{eqn:induction}) changes the number of magnetic-field lines threading a fluid element. Indeed, the origin of the Hall term is the differential motion between the electrons, to which the magnetic-field lines are tied (modulo Ohmic losses), and the drifting ions, which we take to be collisionally well-coupled to the bulk neutral fluid. 

Since the divergences of both the vorticity and the magnetic field are zero, any new vortex and magnetic-field lines that are made must be created as continuous curves that grow out of points or lines where the vorticity and magnetic field respectively vanish. Put simply, just as the effect of the Lorentz force on the vorticity is non-dissipative, so too is the Hall effect on the magnetic field; vorticity and magnetic flux can only be {\em redistributed} by these processes. We now prove that they must be redistributed in a specific way. 

Consider the canonical momentum,
\begin{equation}
\bb{\wp}_{\rm canonical} \equiv m \bigl( \bb{v} + \bb{\Omega} \btimes \bb{r} \bigr) + \frac{e\bb{A}}{c} \frac{n_{\rm e}}{n} ,
\end{equation}
and the associated {\em canonical vorticity},
\begin{equation}\label{eqn:wcan}
\bb{\omega}_{\rm canonical} \equiv \frac{1}{m} \grad \btimes \bb{\wp}_{\rm canonical}
= \bb{\omega} + \frac{e\bb{B}}{mc} \frac{n_{\rm e}}{n} ,
\end{equation}
where $\bb{A}$ is the magnetic vector potential satisfying $\bb{B} = \grad \btimes \bb{A}$. Students of plasma physics will recognize the final term in equation (\ref{eqn:wcan}) as the vectorized Hall frequency $\omega_{\rm H} \equiv (eB/mc) (n_{\rm e}/n)$, at and above which small-wavelength circularly polarized waves with left-handed polarization cannot propagate. Combining equations (\ref{eqn:induction}) and (\ref{eqn:vorticity}), we find that the canonical vorticity satisfies
\begin{equation}
\D{t}{\bb{\omega}_{\rm canonical}} = \grad \btimes \bigl( \bb{v} \btimes \bb{\omega}_{\rm canonical} \bigr) + \nabla^2 \bigl( \nu \bb{\omega} + \eta \bb{\omega}_{\rm H} \bigr) .
\end{equation}
This equation states that, in the absence of dissipative sinks, the canonical vorticity is frozen into the fluid. As a result, the combined number of vortex and magnetic-field lines threading a material surface is conserved; i.e.~the {\em canonical circulation}
\[
\Gamma_{\rm canonical} \equiv \oint_{\mc{C}} \bb{\wp}_{\rm canonical} \bcdot \,{\rm d} \bb{\ell} \quad \left( = \frac{1}{m} \int_{\mc{S}} \bb{\omega}_{\rm canonical} \bcdot \, {\rm d} \bb{S} \right)
\]
around a simple closed contour $\mc{C}$ bounding a material surface $\mc{S}$ is a constant. This is simply Kelvin's (1869) circulation theorem\nocite{kelvin69} generalized for Hall-MHD. An important consequence is that  {\em a local increase in magnetic flux must be accompanied by a local decrease in vorticity flux} and vice versa.

Such behaviour is absent in ideal MHD, in which the magnetic flux is conserved for each fluid element independent of how the vorticity is advected. The difference is due to the fact that, in Hall-MHD, the ion-neutral fluid drifts relative to the field lines and, as such, has its momentum augmented by the magnetic field through which it travels. One may think of this as a consequence of Lenz's law. We refer the reader to the review by \citet{pm01} for further discussion of conserved quantities in Hall-MHD.

 %
 %
\subsection{Hall electric field and Maxwell stress}\label{sec:hallmax}

Many of the results in this Paper stem from the realisation that the Hall electric field may be re-written in the following form:
\begin{equation}\label{eqn:jxb}
\frac{ \bb{J} \times \bb{B} }{ce n_{\rm e}} =  \grad \bcdot \left( \frac{\bb{B} \bb{B}}{4\pi e n_{\rm e}} \right),
\end{equation}
dropping the extra $\grad B^2$ term with impunity. This form is particularly useful, as it underscores the connexion between the evolution of the magnetic flux and the Maxwell stress
\[
M_{ij} \equiv \frac{ B_i B_j }{4\pi} ,
\]
whose $xy$-component plays the dominant role in transporting angular momentum in MRI-driven turbulence. In other words, {\em the transport of magnetic flux in a partially ionised accretion disc is intimately tied to the efficiency and nature of the angular-momentum transport}. This, along with the conservation of canonical vorticity, will turn out to be an extremely important property for understanding the subsequent analytical and numerical results.

%
 %
\subsection{Hall effect in protoplanetary discs}\label{sec:ppds}

The Hall effect becomes important on lengthscales $\lesssim$$\lh$, where
\begin{equation}
\lh \equiv \frac{v_{\rm A}}{\omega_{\rm H}} = \left( \frac{ m_{\rm i} c^2 }{ 4 \pi Z^2 e^2 n_{\rm i} } \right)^{1/2} \left( \frac{ \rho }{ \rho_{\rm i} } \right)^{1/2}
\end{equation}
and $v_{\rm A} \equiv B / ( 4 \pi \rho )^{1/2}$ is the Alfv\'{e}n speed. The lengthscale $\lh$ is the ion skin depth divided by the square root of the mass-weighted ionisation fraction, and is independent of magnetic-field strength. The smaller the degree of ionization, the broader the range of scales that can be appreciably affected. This is why the cold, dense regions of protoplanetary discs are so easily susceptible to the Hall effect.

While the complexity involved in diagnosing the ionization rates, chemical abundances, and consequent diffusivities in actual protoplanetary discs cannot be overstated \citep[for a review, see][]{wardle07}, it helps if we have at least some handle, if only rough, on the importance of the Hall effect in such discs. Assuming an equilibrium balance between cosmic-ray ionization and dissociative recombination leads to the scaling $n_{\rm e} \propto \sqrt{n}$, for which $\lh$ is a constant dependent only upon the ionization and recombination rates and the mean mass per particle. Taking their respective values to be $\zeta_{\rm cr} = 10^{-17}~{\rm cm}^3~{\rm s}^{-1}$, $\alpha_{\rm dr} = 5 \times 10^{-7}~{\rm cm}^3~{\rm s}^{-1}$, and $m = 2.33m_{\rm p}$ \citep[e.g.][]{un90}, we find $\lh \simeq 0.4~{\rm au}$. Assuming the standard model of the minimum-mass solar nebula \citep[MMSN;][]{hayashi81}, the disc scale-height
\[
H \simeq 0.03 \left( \frac{r}{1~{\rm au}} \right)^{5/4} ~{\rm au}
\]
is comparable to $\lh$ at a radius $r \sim 10~{\rm au}$, inside of which the entire disc thickness becomes subject to the Hall effect. Further in around $\sim$$1~{\rm au}$, there is enough column density to effectively shield cosmic rays; there, the Hall effect is likely to be most important away from the midplane where Ohmic losses are less severe.

The presence of a small fraction of dust grains ($\sim$$10^{-4}$--$10^{-2}$ by mass) can affect the extent of the Hall-dominated region by a substantial, though highly uncertain, amount by soaking up gas-phase charges and altering the equilibrium balance of chemical reactions. In this Paper, we circumvent these complications by restricting ourselves to radii $\sim$$5$--$10~{\rm au}$ where both Hall and Ohmic diffusion are considered to be dominant, and by allowing for a range of fields strengths and ionization fractions in a parameter study. Readers interested in the details of protoplanetary-disc chemistry, the consequent values of the diffusivities, and the implications for the (linear) MRI may consult, e.g., \citet{sw08} and \citet{ws12}.

%
%
\subsection{Dimensionless free parameters}\label{sec:params}

Solutions to equations (\ref{eqn:force})--(\ref{eqn:divb}) are governed by four dimensionless free parameters: the plasma beta, 
\[
\beta \equiv \frac{( \Omega H )^2}{\vasq} ;
\]
the viscous Elsasser number,
\[
\Lambda_{\nu} \equiv \frac{ \vasq }{ \nu \Omega } ;
\]
the (classical) Elsasser number,
\[
\Lambda_{\eta} \equiv \frac{ \vasq }{ \eta \Omega } ;
\]
and the Hall Elsasser number,
\[
\Lambda_{\rm H} \equiv \frac{ \omega_{\rm H}}{\Omega} .
\]
The Hall Elsasser number can be cast in the more traditional form,
\[
\Lambda_{\rm H} \equiv \frac{ \vasq }{ \eta_{\rm H} \Omega } ,
\]
by introducing an effective Hall resistivity, $\eta_{\rm H} \equiv \vasq / \omega_{\rm H} = v_{\rm A} \lh$; however, this form is somewhat specious as the Hall effect is not dissipative. We also define the Reynolds number,
\[
{\rm Re} \equiv \frac{ \Omega H^2}{\nu} ,
\]
and the magnetic Reynolds number,
\[
{\rm Rm} \equiv \frac{ \Omega H^2}{\eta} ,
\]
which differ from their Elsasser counterparts by a factor of $\beta$.

The ratio of viscosity to resistivity is known as the magnetic Prandtl number, ${\rm Pm} = \nu / \eta$. While ${\rm Pm}$ does not appear explicitly in the equations, it is known to affect the saturated state of magnetorotational turbulence \citep{ll07,fplh07,sh09,ll10}. In protoplanetary discs, our primary systems of interest here, ${\rm Pm} \ll 1$ and thus we restrict our attention to low-${\rm Pm}$ flows. However, it should be noted that high-${\rm Pm}$ flows could exhibit significantly different behaviour than what is presented here \citep[e.g.][]{bh08}.

Taking the density, temperature, and rotation frequency at a radius of $10~{\rm au}$ in the MMSN, and assuming a $10~{\rm mG}$ magnetic field and $\mu{\rm m}$-sized dust grains \citep[see fig.~1 in][]{sw08}, we find typical values for these parameters of $\beta \approx 1000$, $\Lambda_\eta \approx 0.5$, $\Lambda_{\rm H} \approx 0.01$, $\lh  \approx 3H$, and ${\rm Rm} \approx 500$; for all practical purposes, ${\rm Re}$ is infinite in protoplanetary discs. We adopt similar parameters in our numerical simulations (see Table \ref{tab:runs}) with the exception of ${\rm Re}$; numerical constraints demand that its value be $\lesssim$$10^4$.

%
%
\subsection{Hall-MRI channel modes}\label{sec:channel}

It is well known that, in the absence of the Hall effect, equations (\ref{eqn:force})--(\ref{eqn:divb}) admit exact nonlinear solutions referred to as MRI `channel' modes:
\begin{eqnarray}\label{eqn:bch}
\bb{B} &=& \bb{B}_0 + \bb{B}_{\rm ch}
\nonumber\\*
\mbox{} &=& B_0 \ez  + b e^{\gamma t} B_0 \cos Kz \left( \ex \sin \theta - \ey \cos \theta \right) ,
\end{eqnarray}
\begin{eqnarray}\label{eqn:vch}
\bb{v} &=& \bb{v}_0 + \bb{v}_{\rm ch} 
\nonumber\\*
\mbox{} &=& 2 A x \ey + b e^{\gamma t} v_0 \sin Kz \left( \ex \cos \phi + \ey \sin \phi \right) ,
\end{eqnarray}
where $\gamma$ is the growth rate of the mode, $K$ is its vertical wavenumber, $b$ is a dimensionless measure of the channel amplitude, and $v_0$ and $B_0$ are constants \citep{gx94}. In the absence of dissipation, the two (constant) orientation angles $\phi$ and $\theta$ are equal. These solutions are exact because all nonlinearities vanish (i.e. $\bb{v}_{\rm ch} \bcdot \grad \bb{v}_{\rm ch} = \bb{B}_{\rm ch} \bcdot \grad \bb{B}_{\rm ch} = \bb{v}_{\rm ch} \bcdot \grad \bb{B}_{\rm ch} = \bb{B}_{\rm ch} \bcdot \grad \bb{v}_{\rm ch} = 0$).

What seems to have gone unappreciated in the Hall-MRI literature is that these channel modes remain exact nonlinear solutions in the presence of the Hall effect, since the channel current density
\begin{equation}\label{eqn:jch}
\bb{J}_{\rm ch} = - \frac{K c}{4\pi} \, b e^{\gamma t} B_0 \sin Kz \left( \ex \cos \theta + \ey \sin \theta \right)
\end{equation}
satisfies $\bb{J}_{\rm ch} \bcdot \grad \bb{B}_{\rm ch} = \bb{B}_{\rm ch} \bcdot \grad \bb{J}_{\rm ch} = 0$. This result also follows from equation (\ref{eqn:jxb}), since there are no nonlinear $z$-components of the Maxwell stress for a magnetic field described by equation (\ref{eqn:bch}). 

Defining a dimensionless Hall parameter,\footnote{This definition differs from that used in \citet{bt01} by a factor of 4 and in \citet{kunz08} by a factor of $| \Omega / A |$.}
\[
{\rm Ha} \equiv \frac{K^2 B_0 c}{8 \pi e n_{\rm e} \Omega } = \frac{1}{2} \frac{\Omega}{\omega_{\rm H,0}} \left( \frac{ K v_{\rm A,0} } {\Omega} \right)^2 ,
\]
and an effective magnetic Reynolds number,
\[
{\rm Rm}_{\rm eff} \equiv \frac{\Omega}{\eta K^2} \left( \frac{1 - {\rm Ha} }{1 - {\rm Pm}} \right) ,
\]
the properties of the channel solution may be found after some straightforward but tedious algebra:
\begin{equation}\label{eqn:growthchannel}
\gamma + \nu K^2 = - A \sin 2\theta - \frac{\Omega}{{\rm Rm}_{\rm eff}} \left( 1 + \frac{A}{\Omega} \sin^2\theta \right) ,
\end{equation}
\begin{eqnarray}\label{eqn:kvachannel}
\lefteqn{
\left( K v_{\rm A,0} \right)^2 = -4 \Omega^2 \left( {\rm Ha} + \frac{A}{\Omega} \sin^2\theta \right)
}\\*&&\mbox{}
\times \left[ 1 + \frac{1}{4} {\rm Rm}^{-2}_{\rm eff} + \frac{A}{\Omega} \cos^2\theta \left( 1 + \frac{1}{2} {\rm Rm}^{-1}_{\rm eff} \tan\theta \right)^2 \right] , \nonumber
\end{eqnarray}
\begin{equation}\label{eqn:kv0channel}
K v_0 = -2 \Omega \left( {\rm Ha} + \frac{A}{\Omega} \sin^2\theta \right) \left( 1 + \frac{1}{4} {\rm Rm}^{-2}_{\rm eff}\right)^{1/2} ,
\end{equation}
\begin{equation}\label{eqn:phitheta}
\phi = \theta - \tan^{-1} \left( \frac{1}{2} {\rm Rm}^{-1}_{\rm eff} \right) .
\end{equation}
In the limit ${\rm Rm}_{\rm eff} \rightarrow \infty$, equations (\ref{eqn:kvachannel}) and (\ref{eqn:kv0channel}) differ from equations (6) and (7) of \citet{gx94} only by the multiplicative factor
\[
\left[ 1 + \frac{2\Omega}{\omega_{\rm H,0}} \left( 1 + \frac{A}{\Omega} \cos^2\theta \right) \right]^{-1} ;
\]
the growth rate in this limit, $\gamma = -A \sin 2 \theta$, is identical to that of the ideal MHD case (their eq.~5). In other words, while the Hall effect cannot alter the growth rate of these modes, it can increase or decrease their wavenumbers and Mach numbers when the angular velocity and background magnetic field are anti-parallel or parallel, respectively \citep{wardle99}.\footnote{For $\Omega / \omega_{\rm H,0} \rightarrow +\infty$, the Hall parameter ${\rm Ha} \rightarrow -(A/\Omega) \sin^2\theta$ and both the channel wavenumber (eq.~\ref{eqn:kvachannel}) and the Mach number (eq.~\ref{eqn:kv0channel}) tend to zero as the destabilising shear is marginalised. Exact equality can only be achieved for ${\rm Ha} = \theta = 0$ or ${\rm Ha} = -(A/\Omega) \sin^2\theta = 1$, both of which are spurious solutions not satisfied by the original equations of motion.} Physically, this sensitivity to the polarity of the magnetic field appears because the magnetic `epicycles' induced by the Hall electric field introduce a handedness that enhances (if $\Omega / \omega_{\rm H,0} < 0$) or reduces (if $\Omega / \omega_{\rm H,0} > 0$) the effective magnitude of the destabilising shear. For
\begin{equation}\label{eqn:hmristable}
\frac{\omega_{\rm H,0}}{\Omega} < -2 \left( 1 + \frac{A}{\Omega} \right) \; \bigl( = - \frac{1}{2}~\textrm{for Keplerian rotation} \bigr) ,
\end{equation}
this multiplicative factor becomes negative at all orientation angles, and the channel modes become stable whistler waves. Note that the Hall-dominated regime is not accessible to the MRI when $B_z < 0$.

By differentiating equation (\ref{eqn:growthchannel}) with respect to $\theta$ and setting the result to zero, one may show that the most vigorously growing channel takes an orientation of
\begin{equation}
\theta_{\rm max} = \phi_{\rm max} = \frac{\pi}{4} + \frac{1}{2} \cot^{-1} \left( 2 {\rm Rm}_{\rm eff} \right)
\end{equation}
and exhibits the growth rate
\begin{eqnarray}\label{eqn:smax}
\lefteqn{
\gamma_{\rm max} + \nu K^2 = - A \left( 1 + \frac{1}{4} {\rm Rm}^{-2}_{\rm eff} \right)^{1/2} - \frac{\Omega}{{\rm Rm}_{\rm eff}} \left( 1 + \frac{A}{2\Omega} \right) .
}\nonumber\\*&&\mbox{}
\end{eqnarray}
In the limit of vanishing viscosity and resistivity, this reduces to the well-known result that the maximum growth rate of the MRI is given by the Oort `A' value \citep{bh91}.

Being nonlinear solutions, Hall-MRI channel modes are analytically amenable to a stability analysis. Such an analysis indicates that these channels are subject to secondary `parasitic' instabilities similar to those found by \citet{gx94} and further studied by \citet{pg09}, \citet{llb09}, and \citet{pessah10}. We defer this calculation to Appendix \ref{app:parasite}, as our numerical simulations indicate that they do not play an important role in the saturation of the MRI in the Hall-dominated regime.

%
%
\section{Shearing-box simulations}\label{sec:simulations}

%
%
\subsection{Numerical approach}\label{sec:numerics}

%
%
\subsubsection{The Snoopy code}

Equations (\ref{eqn:force})--(\ref{eqn:divb}) are solved using Snoopy, a three-dimensional (3D) incompressible spectral code. Snoopy is based on the FFTW 3 library to compute 3D Fourier transforms using domain decomposition. It uses a low-storage, third-order Runge-Kutta (RK3) scheme to compute all the terms except for the linear dissipation terms, which are integrated implicitly. All non-linearities are solved using a pseudo-spectral algorithm, avoiding aliasing errors with the 2/3 rule \citep{chqz88}. Shearing-sheet boundary conditions are implemented by solving the equations in a sheared frame comoving with the mean Keplerian flow. This algorithm is similar to the FARGO scheme \citep{masset00}, although it can be extended to arbitrary order in space and time. This procedure implies a periodic remap in Fourier space, which is performed every $\Delta t_{\rm remap} = |2A| L_y / L_x$, where $L_x$ and $L_y$ are respectively the radial and azimuthal extents of the shearing box \citep{ur04}. Snoopy is now a well-tested code, having been used for both hydro- and magnetohydrodynamical  problems such as resistive MRI, dynamo, and subcritical transitions to turbulence.

We have implemented a new module in Snoopy to account for the Hall effect. This scheme integrates the Hall term using the same RK3 and pseudo-spectral algorithms employed elsewhere in the code. To guarantee stability, the integration time step must be smaller than all of the physical timescales of the system. When the Hall effect dominates the dynamics, the shortest physical timescale is given by the whistler-wave timescale at the grid, $\tau_{\rm w} \equiv \omega_{\rm H}^{-1} ( \Delta x / \pi \lh )^2$. This constraint is quite demanding, as it implies that very short time steps are needed to ensure stability ($\Delta t < \sqrt{3} \, \tau_{\rm w}$; see Appendix \ref{app:hallstab}). The numerical results presented in Section \ref{sec:simresults} are therefore obtained at limited resolution (typically 64 points per $H$); even at this resolution, a simulation of the Hall-dominated MRI requires $\sim$$50$ times more computational time than an ideal MHD simulation.

%
%
\subsubsection{Testing the Hall-MHD module}\label{sec:test}

%
%
\begin{figure}
\centering
\includegraphics[width=8cm,clip]{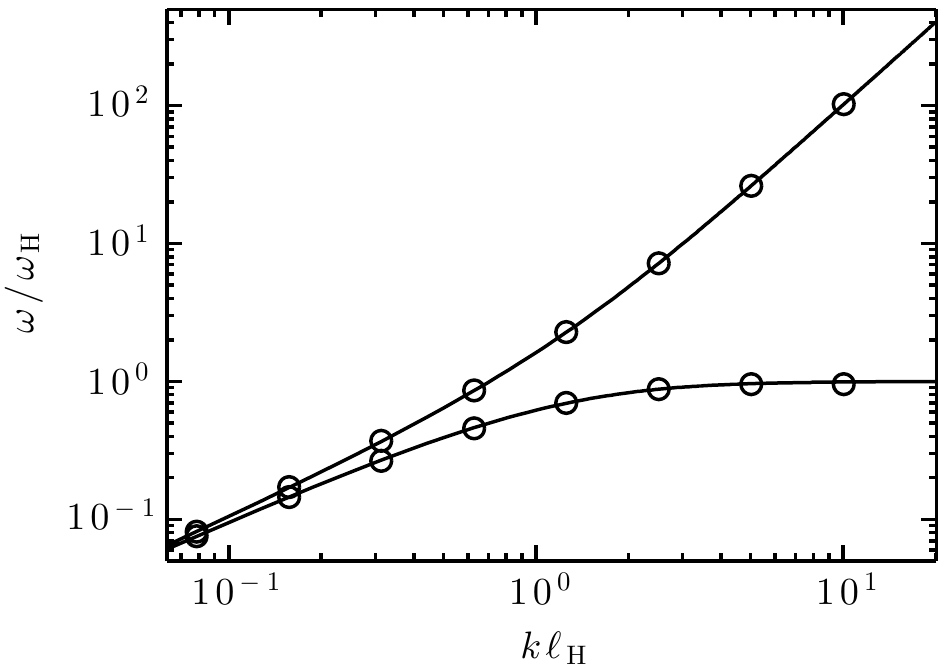}
\caption{Comparison between the analytical dispersion relation (eq.~\ref{eqn:whistlers}; solid lines) and the numerical eigenfrequencies (circles) of linear waves in Snoopy. }
\label{fig:whistlers}
\end{figure}

We have assessed the stability and accuracy of the Hall-MHD module in Snoopy using two tests. First, we verified that linear waves can propagate stably in all three spatial directions while satisfying the linear dispersion relation \citep[eq.~36 of][]{bt01}
\begin{equation}\label{eqn:whistlers}
\frac{\omega}{ \omega_{\rm H}} = k \lh \left[ \sqrt{ 1 + \left( \frac{ k \lh }{ 2 } \right)^2 } \pm \frac{k\lh }{ 2 } \right]
\end{equation}
across a range of wavenumbers $k$. At small wavenumbers (low frequencies), these waves are circularly-polarised Alfv\'{e}n waves; at large wavenumbers, right-handed waves (plus sign) go over to the high-frequency whistler-wave branch, whereas left-handed waves (minus sign) are cut off at $\omega_{\rm H}$. In Figure \ref{fig:whistlers}, the numerical eigenfrequencies (circles) are overlaid on the two solutions (solid lines) of equation (\ref{eqn:whistlers}). The former were obtained by exciting a small-amplitude velocity perturbation $\delta v_x = 2 \times 10^{-5} \cos(kz)$ along a mean magnetic field $B_0 \ez$ and Fourier transforming $v_x(z=0,t)$ in time. This procedure gives two peaks in the spectra, which correspond to the eigenfrequencies of the right- and left-handed waves. The agreement between the analytical and numerical solutions is very good all the way down to the grid scale (the Nyquist frequency $k_{\rm N} \lh = 3.2 \pi$), a benefit of Snoopy's spectral decomposition. Note that these tests were carried out with $\nu = \eta = 0$, demonstrating that the code can stably propagate whistler waves without the need for explicit dissipation. In Appendix \ref{app:hallstab}, we prove analytically why this is true for the RK3 integration scheme used in Snoopy.

%
%
\begin{figure}
\centering
\includegraphics[width=8cm,clip]{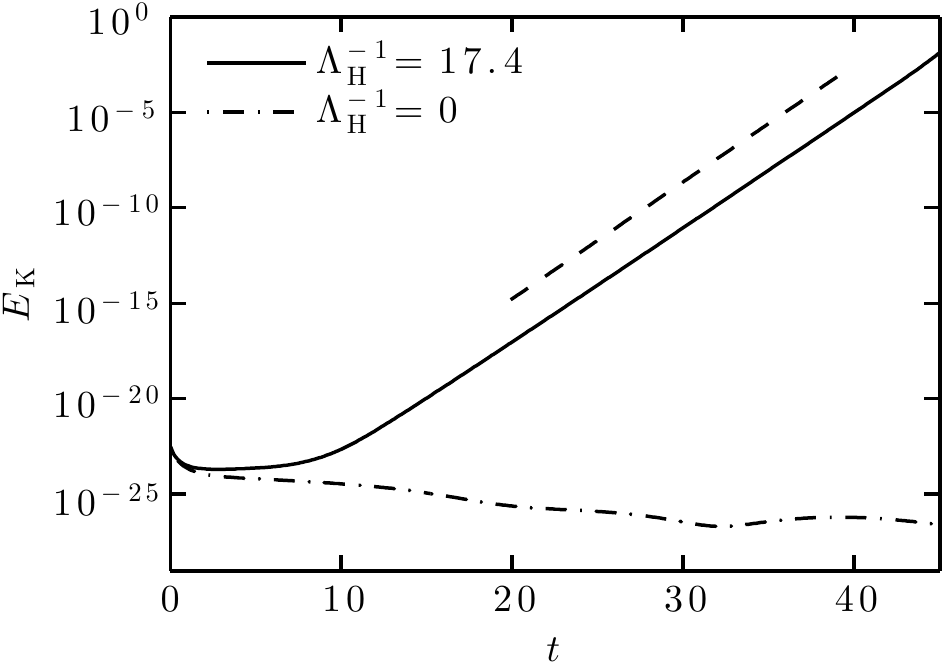}
\caption{Evolution of kinetic energy in a non-rotating shearing box demonstrating the Hall-shear instability. Solid line: with Hall effect and Ohmic dissipation; dot-dashed line: with Ohmic dissipation only. The dashed line represents the theoretical growth rate.}
\label{fig:hallshear}
\end{figure}

For our second test, we simulated the Hall-shear instability described by Kunz (2008; hereafter, K08)\nocite{kunz08}. The basic ingredients of this instability are a background shear, which generates a stream-wise magnetic-field component from a transverse one, and the Hall effect, which induces a circular polarisation that (for certain wavevectors) conservatively reorients stream-wise magnetic fields into the transverse direction. The induced transverse component is sheared further and an exponentially growing instability ensues. Since the interaction between shear and the Hall effect is strongest when the motions implied by the shear lie in the same plane as the magnetic `epicycles' induced by the Hall effect, the instability is maximised when the vorticity, wavevector, and magnetic field all share a mutual axis: $(\bb{k} \bcdot \bb{B} ) ( \bb{k} \bcdot \bb{\omega} )$ must be negative for instability. Note that this instability occurs even in non-rotating systems stable to the `classical' MRI. In Figure \ref{fig:hallshear} we present the results of two numerical experiments with $\Omega = 0$, $A = -3/4$, and $(\Lambda^{-1}_\nu , \Lambda^{-1}_\eta ,  \beta ) = ( 0.2, 1 , 1000)$. We find a linear instability with a growth rate $\gamma = 0.69$ when $\Lambda^{-1}_{\rm H} = 17.4$, matching the theoretical growth rate of the most unstable mode ($\gamma \simeq 0.70$) within 2\%. When $\Lambda^{-1}_{\rm H} = 0$, no instability occurs; instead, the sheared fluctuations resistively decay. These results confirm the accuracy of the Hall-MHD module in Snoopy.

%
%
\begin{table*}
\centering
\caption{\label{tab:runs} List of the runs discussed in this Paper, along with their defining dimensionless parameters. The viscous, resistive, and Hall Elsasser numbers all refer to their values in the initial state, as does the plasma beta parameter; the Hall lengthscale $\lh$ is constant. The growth rate $\gamma$ refers to the most unstable pure-$k_z$ mode available in the simulation domain. The time-averaged turbulent transport $\overline{\alpha}$ is obtained using data between $t=100$ and $630$ (unless otherwise noted). }
\begin{minipage}{0.9\linewidth}
\centering
\renewcommand{\thefootnote}{\thempfootnote}
\begin{tabular}{|c|c|c|c|c|c|c|c|c|c|c|}
\hline
Name 			& $ L_x\times L_y\times L_z$ 	& $n_x \times n_y \times n_z$  & $\Lambda_\nu^{-1}$ 	& $\Lambda_\eta^{-1}$     & $\lhmi$ 			& $\lh$ 		&$\beta$     & $\gamma$ & $\overline{\alpha}$ 	\\
\hline 
\hline
ZB1I1 			& $4\times 4\times 1$		& $256 \times 128 \times 64$ 		& $0.2$		    	     	& $1$			&   $0$          	   	& $0$		&  $1000$		&0.40   &    $3\times 10^{-2}$           \\	
ZB1H1			& $4\times 4\times 1$		& $256 \times 128 \times 64$ 		& $0.2$		    	     	& $1$     			&   $17.4$         	   	& $0.55$		&  $1000$		&0.71   &    $1.7\times 10^{-4}$         \\
ZB1H1L			& $8\times 8\times 1$             	& $512 \times 256 \times 64$ 		& $0.2$		    	     	& $1$      			&   $17.4$         	   	& $0.55$		&  $1000$		&0.71   &    $1.4\times 10^{-4}$         \\

ZTB1H1			& $4\times 4\times 1$	    	& $256 \times 128 \times 64$		& $0.2$				& $1$			&   $17.4$			& $0.55$		&  $1000$ 	&0.67  &	$3.2 \times 10^{-4}$		\\
ZB1H2			& $4\times 4\times 1$	     	& $256 \times 128 \times 64$		& $0.2$				& $1$			&  $4.2$			& $0.13$		&  $1000$		&0.60   &    $1.4 \times 10^{-1}$	\\
ZB1H3			& $4\times 4\times 1$             	& $256 \times 128 \times 64$ 		& $0.2$		      	   	& $1$      			&   $8.6$        	   	& $0.27$		&  $1000$		&0.62   &    $6.8\times 10^{-2}$         \\
ZB1H4			& $4\times 4\times 1$             	& $256 \times 128 \times 64$ 		& $0.2$		    	     	& $1$     			&   $13$         	   	& $0.41$		&  $1000$		&0.68   &    $4.7\times 10^{-4}$		\\
ZB1H5			& $4\times 4\times 1$             	& $256 \times 128 \times 64$ 		& $0.2$		    	     	& $1$     			&   $21.8$         	   	& $0.69$		&  $1000$		&0.70   &    $1.2 \times 10^{-4}$         \\	
ZB1H6			& $4\times 4\times 1$             	& $256 \times 128 \times 64$ 		& $0.2$		    	     	& $1$     			&   $30.4$         	   	& $0.97$		&  $1000$		&0.53   &    $1.4\times 10^{-5}$         \\	
\hline
ZB3I1			& $4\times 4\times 1$             	& $256 \times 128 \times 64$ 		& $0.32$         			& $1$		      	&   $0$           		& $0$		&   $3200$	&0.40   &    $1.9\times 10^{-2}$         \\
ZB3H1			& $4\times 4\times 1$              	& $256 \times 128 \times 64$         	& $0.32$         			& $1$			&   $-1$           		& $0.018$  	&   $3200$	&0.25   &    $6.7\times 10^{-3}$         \\
ZB3H2			& $4\times 4\times 1$             	& $256 \times 128 \times 64$       	& $0.32$         			& $1$      			&   $1$           	  	& $0.018$ 	&   $3200$	&0.50   &    $4.2\times 10^{-2}$        \\	
ZB3H3			& $4\times 4\times 1$             	& $256 \times 128 \times 64$   		& $0.32$         			& $1$      			&   $2$            		& $0.035$ 	&   $3200$	&0.55   &    $6.6\times 10^{-2}$          \\
ZB3H4			& $4\times 4\times 1$             	& $256 \times 128 \times 64$       	& $0.32$         			& $1$      			&   $4$            		& $0.071$ 	&   $3200$	&0.62   &    $1.0\times 10^{-1}$          \\
ZB3H5			& $4\times 4\times 1$             	& $256 \times 128 \times 64$    	& $0.32$         			& $1$      			&   $8$           		& $0.14$  		&   $3200$ 	&0.64  &    $1.1\times 10^{-1}$         \\	
ZB3H6			& $4\times 4\times 1$             	& $256 \times 128 \times 64$          	& $0.32$         			& $1$      			&   $16$         		& $0.28$   	&   $3200$ 	&0.70   &    $1.1\times 10^{-2}$        \\	
ZB3H7			& $4\times 4\times 1$              	& $256 \times 128 \times 64$           	& $0.32$         			& $1$      			&  $32$         		& $0.57$    	&   $3200$	&0.66   &    $4.7\times 10^{-5}$      \\	
ZB3H8			& $4\times 4\times 1$              	& $256 \times 128 \times 64$          	& $0.32$         			& $1$      			&  $56.57$       	 	& $1.0$	 	&   $3200$	&0.74   &    $1.8\times 10^{-5}$      \\  
ZB3H9			& $4\times 4\times 1$		& $256 \times 128 \times 64$           	& $0.32$         			& $1$      			&  $100$ 			& $1.8$           	&   $3200$	&0.53   &    $6.3\times 10^{-7}$      \\  
ZB3I2$^a$		& $4\times 4\times 1$		& $256 \times 128 \times 64$           	& $0.32$         			& $4$      			&  $0$ 			& $0$           	&   $3200$	&0.17   &    $5.1\times 10^{-3}$   \\ 
ZB3H10$^a$		& $4\times 4\times 1$		& $256 \times 128 \times 64$           	& $0.32$         			& $4$      			&  $-1$ 			& $0.018$       	&   $3200$	&0.08   &    $2.7\times 10^{-4}$   \\ 
ZB3H11 			& $4\times 4\times 1$              	& $256 \times 128 \times 64$           	& $0.32$         			& $4$      			&  $1$ 			& $0.018$        	&   $3200$	&0.24   &    $1.3\times 10^{-2}$      \\  
ZB3H12 			& $4\times 4\times 1$              	& $256 \times 128 \times 64$           	& $0.32$         			& $4$      			&  $2$ 			& $0.035$         &   $3200$  	&0.30   &    $2.4\times 10^{-2}$      \\  
ZB3H13 			& $4\times 4\times 1$              	& $256 \times 128 \times 64$          	& $0.32$         			& $4$      			&  $4$ 			& $0.071$       	&   $3200$	&0.39   &    $5.1\times 10^{-2}$      \\  
ZB3H14 			& $4\times 4\times 1$              	& $256 \times 128 \times 64$          	& $0.32$         			& $4$      			&  $8$ 			& $0.14$           	&   $3200$ 	&0.50  &    $9.5\times 10^{-2}$      \\  
ZB3H15 			& $4\times 4\times 1$              	& $256 \times 128 \times 64$           	& $0.32$         			& $4$      			&  $16$ 			& $0.28$           	&   $3200$ 	&0.56  &    $2.0\times 10^{-2}$	   \\ 
ZB3H16 			& $4\times 4\times 1$              	& $256 \times 128 \times 64$           	& $0.32$         			& $4$      			&  $32$ 			& $0.57$           	&   $3200$	&0.62   &    $4.9\times 10^{-4}$      \\
\hline
ZB10I1			& $4\times 4\times 1$             	& $256 \times 128 \times 64$ 		& $0.2$		     	    	& $50$    			&   $0$      	   		& $0$		&   $10000$	&0.00   &    $< 10^{-20}$                     \\
ZB10I2			& $4\times 4\times 1$             	& $256 \times 128 \times 64$ 		& $0.2$		         		& $1$      			&   $0$      	  		& $0$		&   $10000$ 	&0.41  &    $8.7\times 10^{-3}$   \\
ZB10H1$^{ab}$	& $4\times 4\times 1$	     	& $256 \times 128 \times 64$		& $0.2$				& $50$		         &   $25$			& $0.25$		&   $10000$	&0.19   &	    $2.1\times 10^{-1}$	 \\  
ZB10H2			& $4\times 4\times 1$             	& $256 \times 128 \times 64$ 		& $0.2$		       	  	& $50$     			&   $50$          	   	& $0.5$		&   $10000$ 	&0.32  &    $1.8\times 10^{-3}$       \\
ZB10H3			& $4\times 4\times 1$             	& $256 \times 128 \times 64$ 		& $0.2$		       	  	& $50$     			&   $100$          	   	& $1$		&   $10000$	&0.47   &    $6.7\times 10^{-4}$          \\
ZB10H4			& $4\times 4\times 1$             	& $256 \times 128 \times 64$ 		& $0.2$		         		& $50$     			&   $200$          	   	& $2$		&   $10000$  	&0.55 &    $4.8\times 10^{-5}$           \\
ZB10H5			& $4\times 4\times 1$             	& $256 \times 128 \times 64$ 		& $0.2$		        	 	& $50$     			&   $300$          	   	& $3$		&   $10000$ 	&0.41  &    $1.3\times 10^{-5}$              \\
\hline
\end{tabular}
\vspace{-0.2in}
\footnotetext[1]{These runs exhibit a relatively small growth rate, and so the time-averaging procedure is performed between $t=300$ and $630$ in order to eliminate the influence of initial transients.}
\footnotetext[2]{This run shows strong bursts of turbulence associated with the break-up of channel modes. A longer time average should be used to obtain a properly converged $\overline{\alpha}$; consequently, the value of $\overline{\alpha}$ for this run is not used in the discussion.}
\end{minipage}
\end{table*}

%
%
\subsubsection{Units and runs}

For our Hall-MRI simulations, the equations are put in dimensionless form by choosing units natural to the system. The unit of time $\tau = \Omega^{-1}$ is the inverse of the rotation frequency. Lengthscales are measured in units of the vertical box size $L_z$, and the magnetic-field strength is measured in units of the initial Alfv\'en speed $v_{\rm A,0}$. We also introduce a dimensionless measure of the turbulent transport,
\[
\alpha = \frac{\delta v_x \delta v_y - \delta B_x \delta B_y}{L_z^2\tau^{-2}}.
\]
This definition is made equivalent to the \cite{ss73} definition of $\alpha$ by setting $L_z = H$, which we do for all of our simulations. Keplerian rotation (i.e.~$A = -3/4$) is assumed.

In this work, we only consider horizontally extended (`slab') shearing boxes with $L_x, L_y > L_z$. As discussed by \cite{bmc08}, a slab configuration exhibits better convergence properties than bar configurations (as used by SS02) due to the absence of recurrent channel modes. Horizontally extended domains have also been shown to impact quantitative measurements; \cite{bmc08} reported that the time-averaged turbulent transport in a $L_x=8$ slab is a factor $\sim$$2$ smaller than the transport measured in a $L_x=1$ bar. One should bear this in mind when comparing our results with those of SS02. For most of our simulations, we consider boxes with aspect ratio $L_x\times L_y \times L_z=4\times 4 \times 1$ and resolution $n_x\times n_y \times n_z=256 \times 128 \times 64$; we have also performed two versions of our fiducial simulation, one with an extended domain ($8 \times 8 \times 1$) and one with increased resolution ($384 \times 192 \times 96$).

All the runs discussed in this Paper are summarised in Table \ref{tab:runs}. Unless otherwise stated, the initial conditions are white noise\footnote{The nonlinear outcome of a simulation does not depend significantly on the initial conditions.} on all components of the velocity and magnetic fields with a typical RMS amplitude equal to $7\times10^{-7}$. We add to the magnetic field a constant poloidal component whose magnitude is determined by a choice of initial $\beta$. These runs have been integrated up to $t=630$, corresponding to 100 orbits in a Keplerian shearing box; our fiducial run was integrated to $t=2000$. Finally, we introduce an averaging procedure $\overline{\ \ \ \cdot \ \ \ }$, which denotes a volume and time average. The time average is performed using data only from $t=100$ (or, in some cases, $t=300$) to the end of the run, in order to avoid transient effects from the (arbitrary) initial conditions and the initial growth of channel modes.

We close this Section by providing a translation between our dimensionless parameters and those employed by SS02. Those authors used $X = 2 \lhmi$ and ${\rm Re}_{M} = \Lambda_{\eta}$, and focused mainly on initial values of $X = 1$, $2$, and $4$. For their fiducial initial $\beta = 3200$, this implies $\lh / H \sim \textrm{a few} \times 10^{-2}$. Here we are concerned with somewhat larger values of $X$, which place $\lh \sim H$ and render the entire disc prone to the Hall effect.

%
%
\subsection{Results}\label{sec:simresults}

%
%
\subsubsection{Averaged turbulent transport and the Hall effect}

\begin{figure}
\centering
\includegraphics[width=8cm,clip]{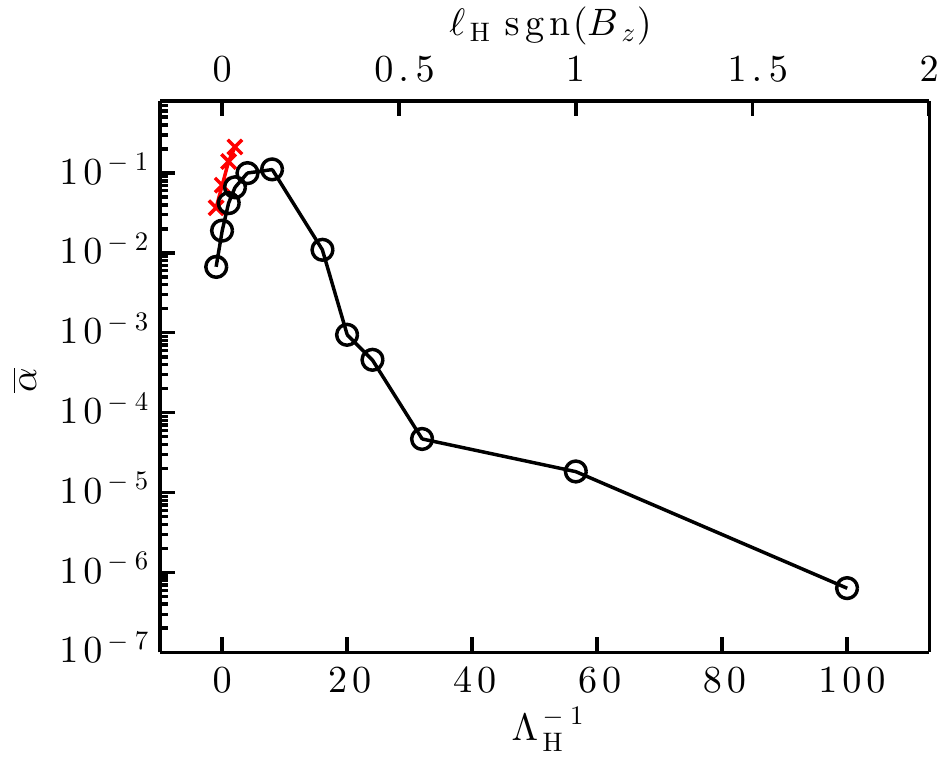}
\caption{Volume-averaged turbulent transport $\overline{\alpha}$ as a function of the inverse Hall Elsasser number $\Lambda^{-1}_\mathrm{H}$ from runs ZB3(I1,\,H1--H9; black circles) and runs Z6--Z9 from SS02 (red crosses).}
\label{fig:alpha-lambda}
\end{figure}

The relationship between the averaged turbulent transport and the intensity of the Hall effect was first explored numerically by SS02. In order to make a proper comparison with this earlier work, we have reproduced runs Z6--Z9 from SS02 (named ZB3H1--ZB3H4 here). We have also explored the regime of very strong Hall effect, up to $\lhmi=100$ (run ZB3H8), corresponding to a Hall term 16 times larger than the largest one considered by SS02. The results of this first exploration are presented in Figure \ref{fig:alpha-lambda}. 

In the regime of weak Hall effect ($\lhmi\lesssim 5$), we qualitatively recover the SS02 results: increasing $\lhmi$ increases the amount of turbulent transport. This demonstrates the sensitivity of disc stability to the orientation of the background magnetic field \citep{wardle99}. The actual values of $\alpha$ we find, however, are significantly smaller than those found by SS02. This is probably due to the different aspect ratios of the shearing boxes employed. 

%
%

\begin{figure}
\centering
\includegraphics[width=8cm,clip]{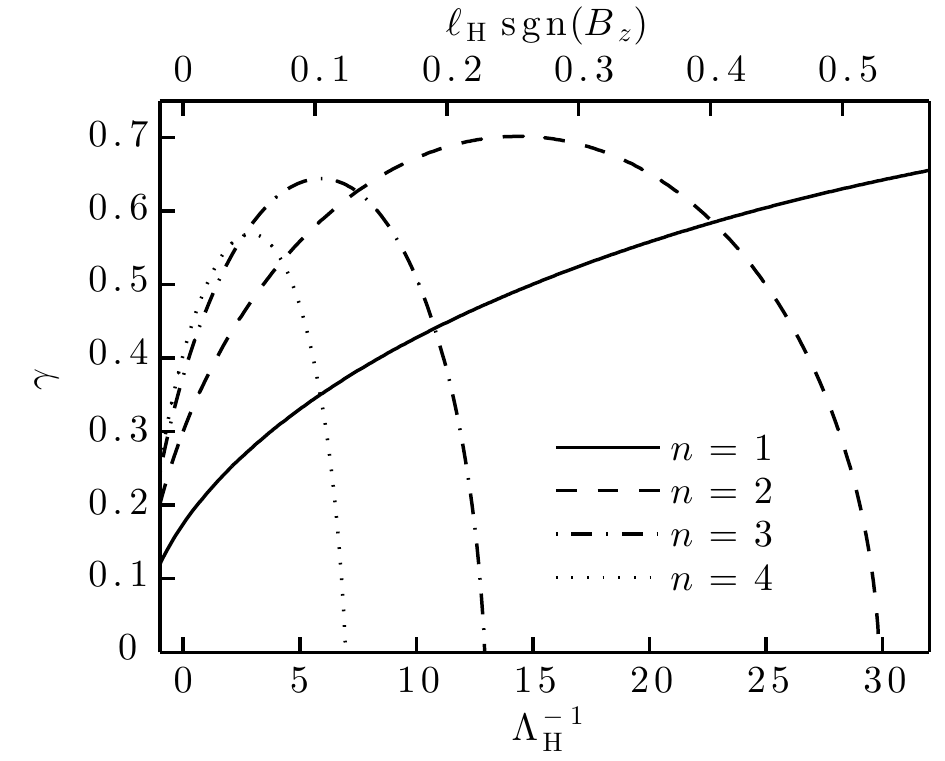}
\caption{Linear growth rate $\gamma$ as a function of the inverse Hall Elsasser number $\lhmi$ for simulations ZB3H(1--9). Each line corresponds to a channel mode with $k_z=2\pi n/L_z$.}
\label{fig:grbeta32}
\end{figure}

For larger $\lhmi$, a totally new behaviour appears: the transport follows a steep decline down to $\overline{\alpha} = 4.7\times 10^{-5}$ at $\lhmi=32$. At first sight, one may think that this behaviour is due to the stabilisation of the MRI when the Hall effect is too dominant. To check this, we present in Figure \ref{fig:grbeta32} the growth rate of the largest channel modes for the parameters used in runs ZB3H(1--4). In each of these runs there exists at least one channel mode with $\gamma > 0.5$; i.e.~a vigorous linear instability is initially present in all of these runs. The behaviour for $\lhmi>5$ indicates that a new saturation mechanism is at work, one that is not related to the linear properties of the flow.

%
%

%
%
\begin{figure}
\centering
\includegraphics[width=8cm,clip]{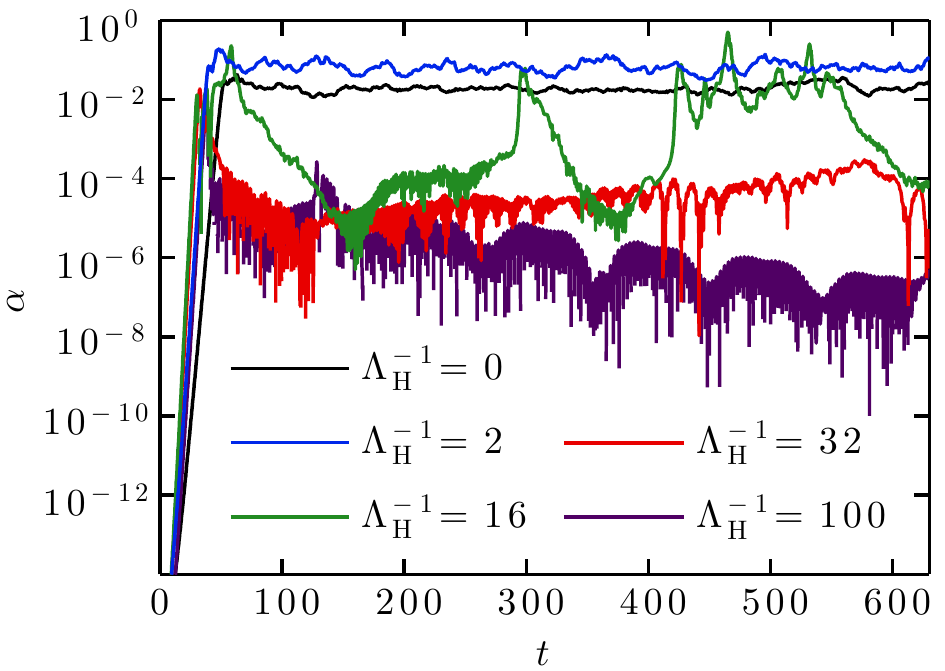}
\caption{Volume-averaged turbulent transport $\alpha$ as a function of time for different inverse Hall Elsasser numbers $\lhmi$ (runs ZB3I1, ZB3H3, ZB3H6, ZB3H7, ZB3H9).}
\label{fig:alpha-time}
\end{figure}

In order to isolate the cause of this new saturation mechanism, we focus on the evolution of the box-averaged turbulent transport at different $\lhmi$ (Fig.~\ref{fig:alpha-time}). Simulations with $\lhmi\lesssim 5$ show the same qualitative behaviour exemplified in ideal- and resistive-MHD simulations of the MRI---a time-steady $\alpha \sim {\rm few} \times 10^{-2}$ with moderate transport spikes. The intermediate case ($\lhmi=16$) shows oscillations between a low-transport state ($\mathrm{LTS}\equiv \alpha<10^{-4}$) and a high-transport state (HTS) qualitatively similar to that seen in the $\lhmi\lesssim 5$ runs. For $\lhmi=32$ and $100$, the turbulence stays in the LTS after an initial burst caused by the breakup of the channel mode. Note that the qualitative behaviour observed during the linear phase in all of these simulations (i.e.~$t\lesssim 50$) is very similar.

\begin{figure*}
\centering
\includegraphics[width=0.90\linewidth,clip]{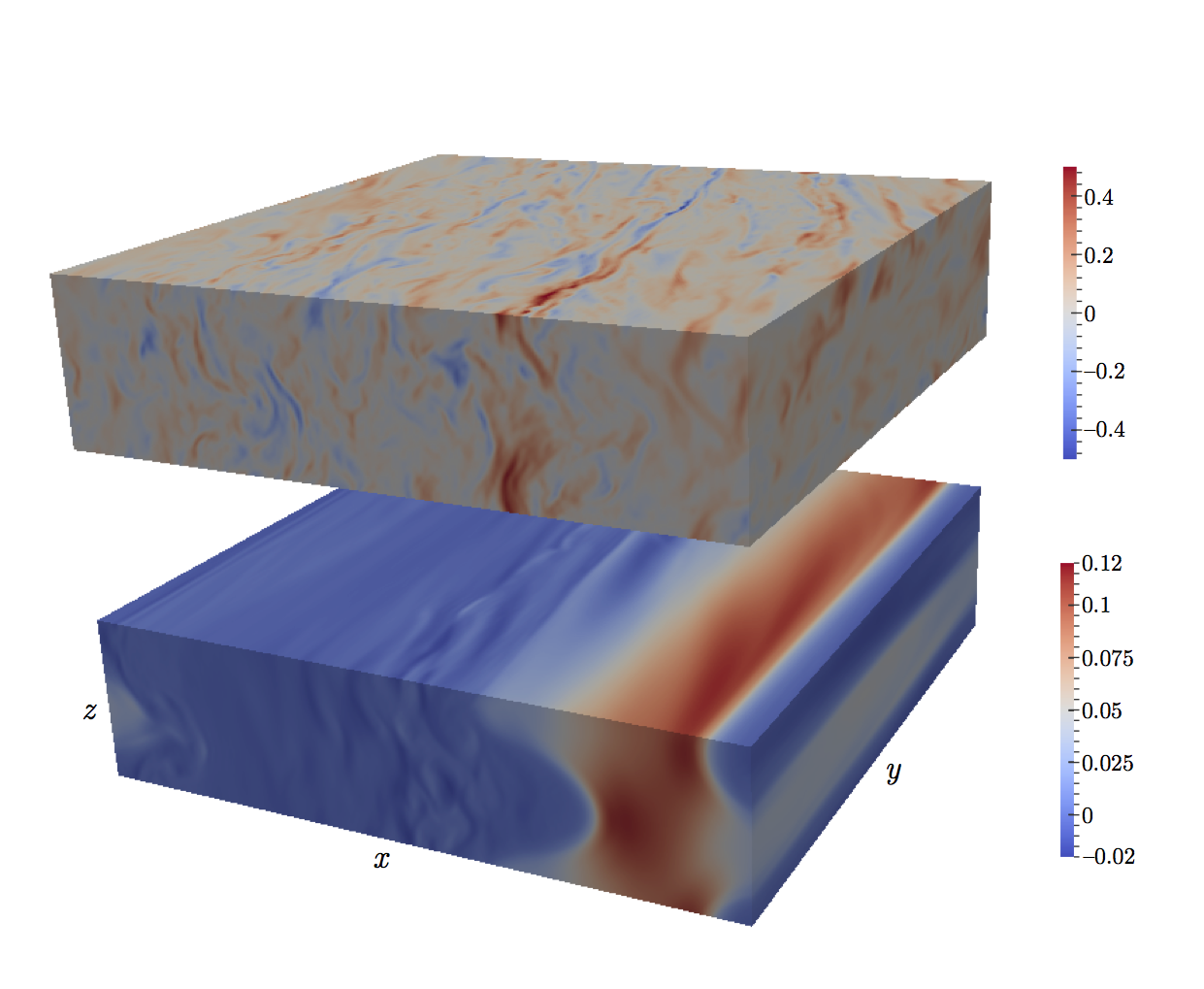}
\caption{Snapshot of the vertical magnetic field $B_z$ in runs ZB3I1 ({\it top}) and ZB3H7 ({\it bottom}) at $t=630$.}
\label{fig:snapshot}
\end{figure*}

Inspecting snapshots of the two extreme cases (Fig.~\ref{fig:snapshot}) reveals the origin of the LTS. In the ideal case ($\lhmi = 0$; top), the flow exhibits turbulent fluctuations of $B_z$. This is the `traditional' saturated state of the MRI as described by \cite{hgb95} and others. In the Hall-dominated case ($\lhmi = 32$; bottom), we observe a coherent, axisymmetric, large-scale structure in $B_z$, which we refer to as a {\em zonal magnetic field}. In this zonal-field configuration, the vertical magnetic flux is accumulated in some radial ($x$) region, leaving most of the box with a very weak $B_z$ (typically $|B_z|<10^{-3}$). In this configuration almost no turbulent activity is observed. Note that the total vertical magnetic flux is conserved, indicating that this feature is due to a \emph{redistribution} of magnetic flux. 

%
%

%
%
\subsubsection{Characterising the low-transport state}\label{sec:lts}

In the previous Section, we described a new saturated state of the MRI called the low-transport state (LTS). This state is characterised by very weak turbulent transport, despite the presence of a vigorous linear instability in the initial equilibrium. Here we conduct a dedicated study of the LTS by examining our fiducial run ZB1H1 in detail. This simulation exhibits the same kind of LTS as described above, with an averaged turbulent transport $\overline{\alpha}=1.8\times 10^{-4}$. Increasing the resolution to $384\times 192\times 96$ does not change the outcome of the saturated state: a similar LTS is observed, demonstrating that our simulations have converged.\footnote{Due to the extreme cost of such a high-resolution simulation, this particular run was stopped at $t=160$; it is therefore not listed in Table \ref{tab:runs}.}

\begin{figure}
\centering
\includegraphics[width=8cm,clip]{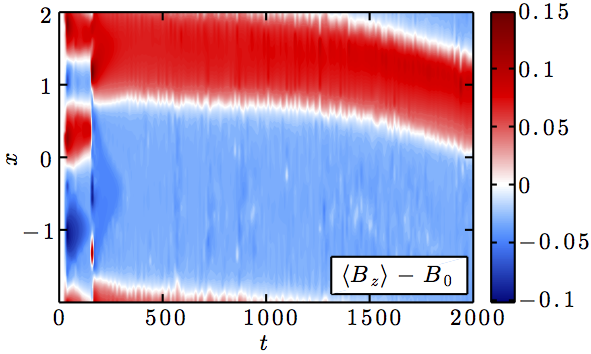}
\newline\newline
\includegraphics[width=8cm,clip]{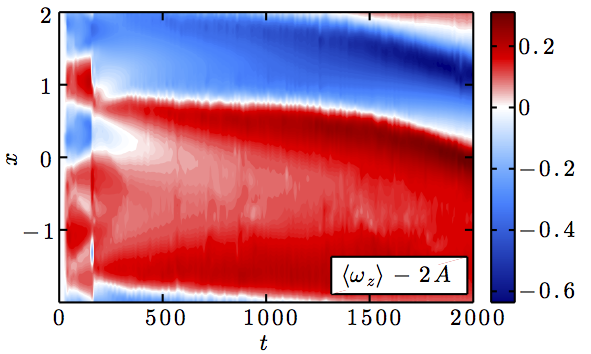}
\newline
\caption{Space-time diagram of the vertically and azimuthally averaged vertical component of the magnetic field $\langle B_z\rangle$ and the vorticity $\langle \omega_z \rangle$ in run ZB1H1. The appearance of vorticity bands, anti-correlated with the zonal-field structures, is observed.}
\label{fig:st-zb1h1}
\end{figure}

We quantify the presence of a zonal field by defining an averaging procedure,
\[
\langle \cdot \rangle \equiv \frac{1}{L_y L_z} \int\!\!\!\! \int {\rm d}y \, {\rm d}z ,
\]
and computing the evolution of the vertically and azimuthally averaged vertical component of the magnetic field $\langle B_z \rangle$. In the top panel of Figure \ref{fig:st-zb1h1} we present the resulting space-time diagram. This diagram clearly exhibits a strong zonal field with a typical radial thickness $\sim$$1$. Outside of this zonal-field region, the averaged field is weak with $|\langle B_z\rangle| \lesssim 10^{-2}$. A closer inspection shows that the system initially exhibits two zonal-field regions, centred at $x\simeq 0.3$ and $x\simeq 1.8$. At $t\simeq 160$ a rapid reorganisation occurs, and these two regions merge to produce one zonal field that survives for more than $1000\,\Omega^{-1}$. While zonal fields are generally very long-lived structures in isolation, this demonstrates that they may become strongly unstable when another is nearby.

%
%

%
%
\begin{figure}
\centering
\includegraphics[width=8cm,clip]{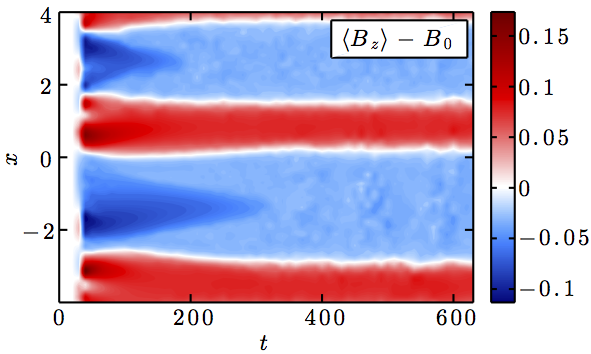}
\caption{Space-time diagram of the vertically and azimuthally averaged vertical component of the magnetic field $\langle B_z\rangle$ in run ZB1H1L. Two stable zonal-field regions are produced.}
\label{fig:st-zb1h1l}
\end{figure}

It was shown in Section \ref{sec:hall} that, in the presence of the Hall effect, a new conserved quantity replaces the magnetic flux: the canonical vorticity. Since canonical vorticity is conserved without dissipative effects (e.g.~viscosity, resistivity), we expect the formation of axisymmetric bands of vertical vorticity that are anti-correlated with the zonal fields. To check for this effect, we have computed the vertically and azimuthally averaged vertical component of the flow vorticity $\langle \omega_z\rangle = -\partial_x \langle v_y \rangle + 2A$ in run ZB1H1. The resulting space-time diagram is shown in Figure \ref{fig:st-zb1h1}b, and clearly demonstrates the formation of a zonal-vorticity region akin to a zonal flow. In accordance with expectations from conservation of canonical vorticity, we also find that the flow vorticity is anti-correlated with the vertical magnetic field. However, the vorticity and magnetic field do not have exactly the same shape---the mean vorticity appears to be concentrated around the edges of the zonal-field region. This difference is due to the explicit dissipation, and in particular to the fact that ${\rm Pm}\ll 1$: magnetic-field lines diffuse quite rapidly through the bulk ion/neutral fluid, whereas vortex lines does not. Therefore, conservation of canonical vorticity is only approximately verified in our simulations, owing to the presence of non-negligible dissipative terms.

Since our box size is limited, one may suspect that the presence of only one zonal-field region in run ZB1H1 is an artifact of the boundary conditions. To check this, we have performed a simulation in a wider box $(8\times 8\times 1)$ with the same physical parameters as run ZB1H1. The space-time diagram of this simulation (run ZB1H1L) is presented in Figure \ref{fig:st-zb1h1l}. We observe the formation of two zonal-field regions of size $\approx$$1.5$, which survive for the remainder of the simulation. This indicates that zonal-field regions have an intrinsic width independent of the radial and azimuthal box size (provided the latter is significantly larger than $H$).

In order to verify that the formation of a zonal field is not dependent upon the initial conditions of the simulation, we have also run a purely resistive ($\lhmi=0$) MRI simulation with the same parameters as run ZB1H1. This simulation was run up until $t=630$, at which point the Hall effect was switched on with $\lhmi=17.4$. Within $\sim$$4$ orbits, the fully developed 3D turbulence disappeared and was replaced by a large-scale zonal field with an averaged turbulent transport $\overline{\alpha} \sim 10^{-4}$. This demonstrates that the LTS, and the zonal field associated with it, are robust nonlinear features of the Hall-dominated MRI.

\begin{figure}
\centering
\includegraphics[width=8cm,clip]{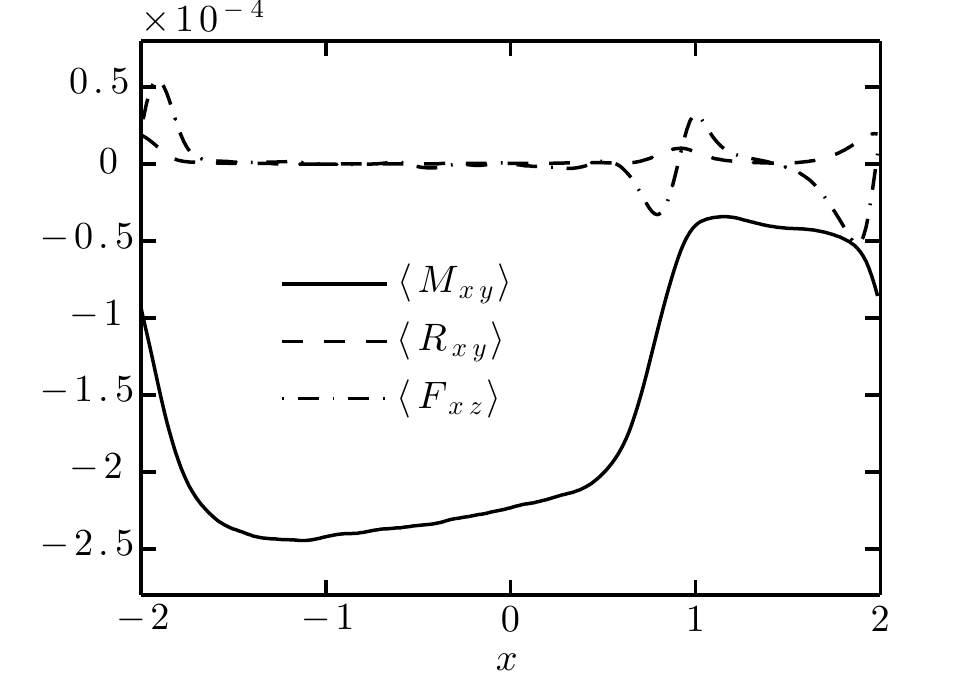}
\newline
\includegraphics[width=8cm,clip]{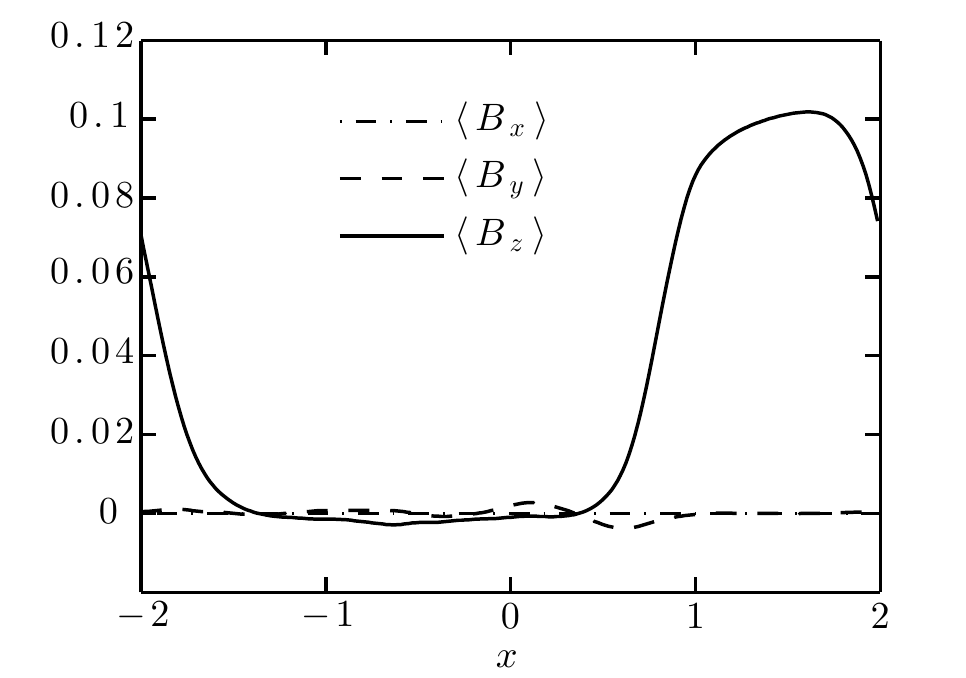}
\newline
\caption{$(y,z,t)$ averages of the turbulent stress ({\it top}) and magnetic field ({\it bottom}) for run ZB1H1. A clear correlation between the Maxwell stress and the mean vertical magnetic field is exhibited, a feature which is responsible for the formation and sustainment of a zonal magnetic field (see Section \ref{sec:lts}).}
\label{fig:profilefid}
\end{figure}

To understand how this zonal-field structure is sustained, we return to the argument given in Section \ref{sec:hallmax}. In particular, it was shown that the Maxwell stress directly enters into the induction equation through the Hall effect. We therefore compute the mean stress and magnetic field in run ZB1H1, averaging these quantities in $y$, $z$, and time (from $t=500$ to $t=600$). The resulting profiles are presented in Figure \ref{fig:profilefid}, and exhibit a very clear correlation between the averaged Maxwell stress $\langle M_{xy} \rangle$ and the magnetic-field profile $\langle B_z \rangle$. In particular, there are inflection points in the Maxwell stress at the boundaries of the zonal-field region. Between these points where the magnetic field is relatively strong, the MRI is magnetically quenched; elsewhere, the field is relatively weak and turbulent fluctuations persist with a low level of transport. As we show explicitly in Section \ref{sec:bifurcation}, such residual fluctuations are responsible for maintaining the integrity of the zonal-field structure in the face of resistive (both molecular and turbulent) diffusion.

%
%

%
%
\subsubsection{A numerical criterion for the low-transport state}

\begin{figure}
\centering
\includegraphics[width=8cm,clip]{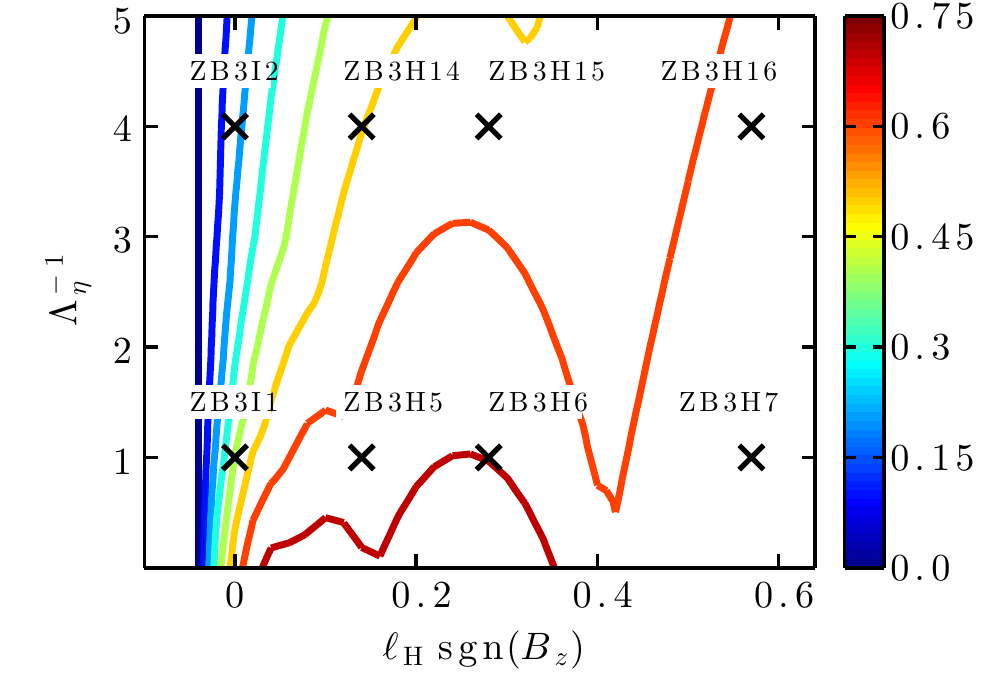}
\newline\newline
\includegraphics[width=8cm,clip]{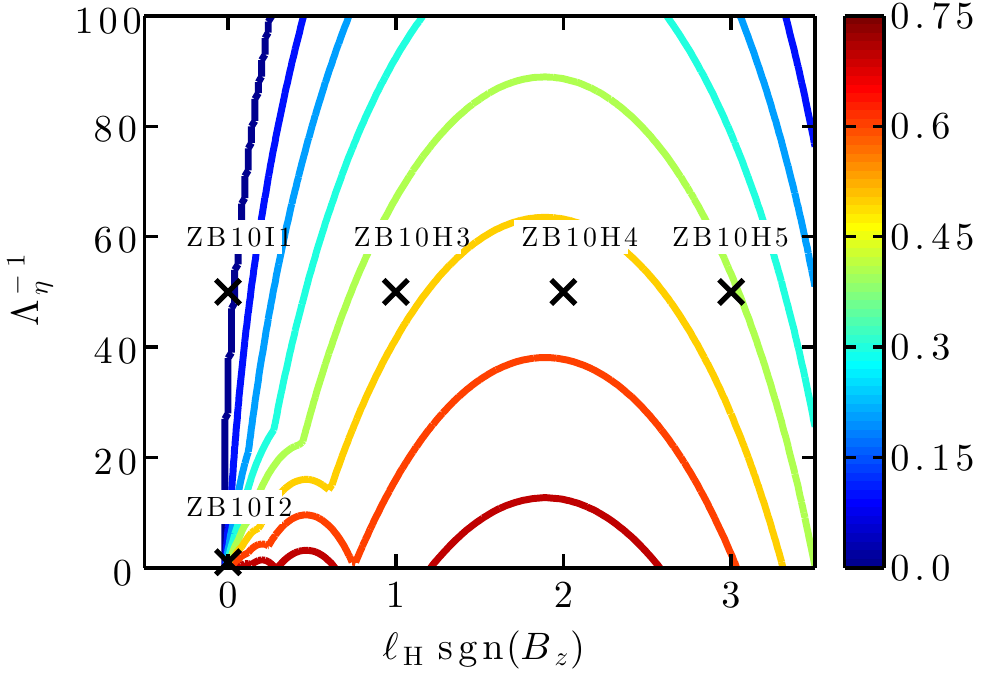}
\newline
\caption{Location of some representative runs from ZB3XX ({\it top}) and ZB10XX ({\it bottom}) on a linear stability diagram. The colour contours denote the growth rates of the most unstable eigenmodes.}
\label{fig:linruns}
\end{figure}

We have shown that an LTS exhibiting axisymmetric (`zonal') fields emerges in several simulations of the Hall-dominated MRI. To make any prediction about the saturation level, one must know when the system will choose the LTS instead of the `classical' turbulent MRI state (HTS). To this end, we have systematically explored the parameter space $(\Lambda_\eta,\,\Lambda_\nu,\,\beta)$. In this parameter space, we include a regime which is stable without the Hall effect (runs ZB10XX). We present in Figure \ref{fig:linruns} the growth rates of the most unstable Hall-MRI modes present in some representative runs from ZB3XX and ZB10XX. As expected, all the runs but ZB10I1 are linearly unstable with growth rates $\gamma > 0.1$. Note that the vertical wavelength of the most unstable mode increases with $\lh$ (\S\,\ref{sec:channel}).

%
%

\begin{figure}
\centering
\includegraphics[width=8cm,clip]{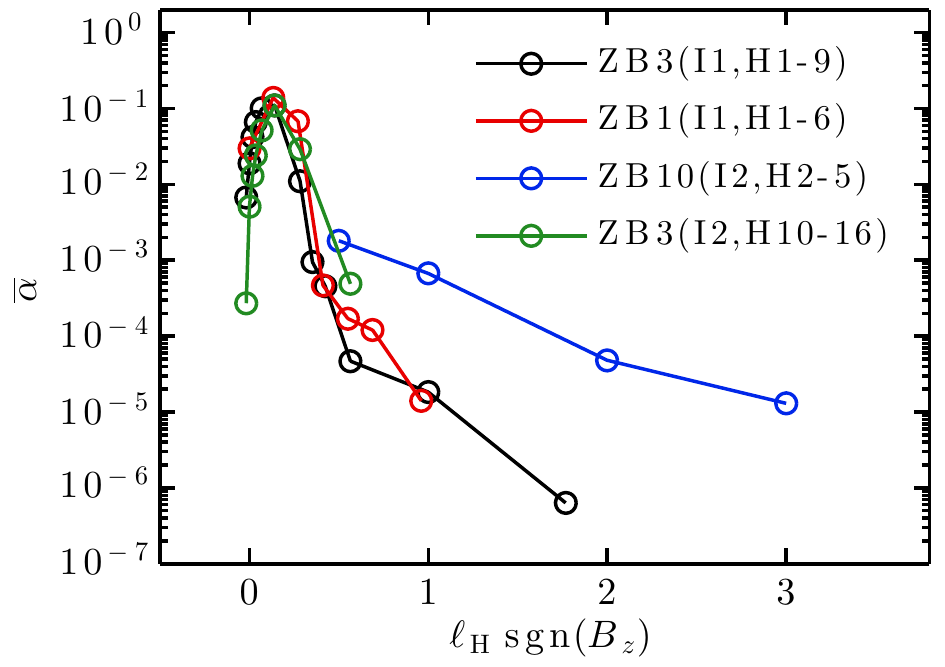}
\caption{Mean turbulent stress versus Hall effect for $\beta=1000$, $3200$, and $10000$, and $\Lambda^{-1}_\eta = 1$, $4$, and $50$. The turbulent stress decreases sharply when $\lh\gtrsim 0.2$, separating the low-transport state from the classical high-transport state.}
\label{fig:alpha-lambdaH}
\end{figure}

In Figure \ref{fig:alpha-lambdaH} we summarise all our results on a single plot exhibiting the mean turbulent stress $\overline{\alpha}$ as a function of $\lh$. Despite the differing initial $\beta$, viscosities, and resistivities, all of the values of $\overline{\alpha}$ tend to collapse onto a single curve dependent primarily upon $\lh$. We find that the system stays in the HTS up to $\lh \simeq 0.2$ for all our simulations, independent of the mean field strength and of the resistivity; in this case the typical turbulent stress $\overline{\alpha}\sim10^{-2}$--$10^{-1}$. Beyond $\lh \sim 0.2$, the system transitions rapidly to an LTS characterised by $\alpha \lesssim 10^{-3}$. A very important characteristic is that the presence of the LTS is not correlated to any linear property of the Hall-MRI. For instance, all of the situations where we observe an LTS are characterised by strongly unstable MRI modes with $\gamma\gtrsim 0.5$ (compare Figs~\ref{fig:linruns} and \ref{fig:alpha-lambdaH}). This illustrates the fact that \emph{a flow which is strongly unstable to the Hall-MRI does not necessarily evolve into fully developed turbulence}.

%
%

%
%
\subsubsection{Other magnetic-field configurations}

One may suspect that the behaviour described above is very specific to situations with only a mean vertical magnetic field. We have therefore run a simulation with a magnetic field inclined by $45^\circ$ in the azimuthal ($y$) direction of strength $\beta=10^3$ (run ZTB1H1). We find results very similar to the pure-$B_z$ run ZB1H1. The transport is increased by a factor $\sim$$2$, but the large-scale zonal-field structure characteristic of the LTS remains.

We have also performed a limited number of simulations either without a mean magnetic field or with only a mean azimuthal magnetic field, finding no sustained zonal-field configurations in either case. While the mean-field theory presented in the next Section does suggest an explanation for this negative result, a more systematic exploration of these cases is needed to verify our conclusions. This is deferred to a future publication.

%
%
\section{Mean-field theory of zonal fields and flows}\label{sec:bifurcation}

In this Section, we formulate a mean-field theory that explains the observed transport bifurcation from a high- to a low-transport state and the appearance of zonal magnetic fields and flows. We start by separating the velocity and magnetic fields into fluctuating and non-fluctuating parts:
\[
\bb{v} = \bb{v}_0 + \langle \bb{v} \rangle + \delta \bb{v} \quad {\rm and} \quad \bb{B} = \langle \bb{B} \rangle + \delta \bb{B} . 
\]
Upon averaging over azimuth and height, equations (\ref{eqn:divv}) and (\ref{eqn:divb}) become $\partial_x \langle v_x \rangle = 0$ and $\partial_x \langle B_x \rangle = 0$, respectively; it follows from equation (\ref{eqn:induction}) that $\langle B_x \rangle = 0$ if it is so initially. For clarity of presentation, we further assume that $\langle v_x \rangle = 0$. Not only is this assumption supported by our numerical results, but it also allows us to cancel global epicyclic oscillations nonessential for understanding the emergence of zonal fields and flows.

Introducing the Reynolds stress,
\[
R_{ij} \equiv \rho \delta v_i \delta v_j ,
\]
and the Faraday tensor,
\[
F_{ij} \equiv \delta v_i \delta B_j - \delta v_j \delta B_i ,
\]
the pertinent mean-field equations are the $z$-component of the averaged induction equation (\ref{eqn:induction}),
\begin{equation}\label{eqn:bzavg}
\D{t}{\langle B_z \rangle} = - \D{x}{\langle F_{xz} \rangle} - \frac{c}{e n_{\rm e}} \DD{x}{ \langle M_{xy} \rangle } + \eta \DD{x}{\langle B_z \rangle} ,
\end{equation}
and the $z$-component of the averaged vorticity equation (\ref{eqn:vorticity}),
\begin{equation}\label{eqn:wzavg}
\D{t}{\langle \omega_z \rangle} = - \frac{1}{\rho} \DD{x}{ \langle R_{xy} \rangle} + \frac{1}{\rho} \DD{x}{ \langle M_{xy} \rangle } + \nu \DD{x}{\langle \omega_z \rangle} .
\end{equation}
The mean vorticity,
\[
\langle \omega_z \rangle =  \D{x}{\langle v_y \rangle} + 2A ,
\]in
includes a contribution from the background shear ($=2A$, which is negative in Keplerian discs). Note that the averaged Maxwell stress $\langle M_{ij} \rangle$ comprises products of only the fluctuating magnetic fields.

In order to solve equations (\ref{eqn:bzavg}) and (\ref{eqn:wzavg}), we must construct models for $\langle F_{ij} \rangle$, $\langle R_{ij} \rangle$, and $\langle M_{ij} \rangle$. The Faraday tensor has been shown to be accurately modeled by a turbulent resistivity with coefficient $\eta_{\rm t}$ \citep{ll09}, and we take this to be the case in what follows. We follow a similar approach for the Reynolds stress by modeling it as a turbulent viscosity with coefficient $\nu_{\rm t}$. Adopting these simplifications, our mean-field equations become
\begin{equation}\label{eqn:bzavg2}
\D{t}{\langle B_z \rangle} \simeq ( \eta + \eta_{\rm t} ) \DD{x}{\langle B_z \rangle} - \frac{c}{e n_{\rm e}} \DD{x}{M} ,
\end{equation}
\begin{equation}\label{eqn:wzavg2}
\D{t}{\langle \omega_z \rangle} \simeq ( \nu + \nu_{\rm t} ) \DD{x}{\langle \omega_z \rangle} + \frac{1}{\rho} \DD{x}{M}  .
\end{equation}
We consider two models for the Maxwell stress, each of which will produce zonal behaviour very similar to that seen in our nonlinear numerical simulations.

%
%
\subsection{Case I: $\langle M_{xy} \rangle = M( \langle B_z \rangle )$}\label{sec:case1}

As a first approach, we take the $xy$-component of the Maxwell stress to be a function only of the local vertical magnetic flux,
\[
\langle M_{xy} \rangle \equiv M( \langle B_z \rangle) ,
\]
and we concentrate on the evolutionary evolution for the mean vertical magnetic field (eq.~\ref{eqn:bzavg2}). We suppose that there is some $\langle B_z^0 \rangle$ that satisfies this equation in steady-state (e.g.~$\langle B^0_z \rangle$ constant) and we examine small deviations $\langle B_z^1\rangle$ about that state. Linearising equation (\ref{eqn:bzavg2}), we find that such deviations satisfy
\begin{equation}\label{eqn:dbzavg}
\D{t}{\langle B_z^1 \rangle} \simeq  \left( \eta+\eta_{\rm t} - \frac{c}{e n_{\rm e}} \left. \deriv{\langle B_z \rangle}{M} \right|_{\langle B^0_z \rangle} \right) \DD{x}{\langle B_z^1 \rangle} .
\end{equation}
This equation has a simple interpretation. While resistivity acts diffusively on $\langle B^1_z \rangle$, the Hall term may be diffusive or anti-diffusive depending upon the local gradient of the Maxwell stress. 

Fortunately, even without a specific model for $M$, progress can be made. For sufficiently large values of $\langle B_z \rangle$, we expect the MRI to be stable and $M\rightarrow 0$. We also expect $M\rightarrow 0$ for sufficiently small values of $\langle B_z \rangle$, since unstable modes exist only at small wavelengths where Ohmic dissipation becomes important and suppresses turbulent transport. In between these extremes, we know that the Maxwell stress is negative since the MRI transports angular momentum outwards. Therefore, ${\rm d}^2 M / {\rm d} \langle B_z \rangle^2 < 0$, and so there must be a value of $\langle B^0_z \rangle = B_{z{\rm ,crit}}$ above which ${\rm d} M / {\rm d} \langle B_z \rangle > 0$ and below which ${\rm d} M / {\rm d} \langle B_z \rangle < 0$.

First, let us consider $\langle B^0_z \rangle < B_{z{\rm ,crit}}$. Then both the Ohmic and Hall contributions to equation (\ref{eqn:dbzavg}) are positive and any deviations from steady-state diffusively decay. Now let us consider the opposite case, $\langle B^0_z \rangle > B_{z{\rm ,crit}}$. Then the Ohmic and Hall contributions to equation (\ref{eqn:dbzavg}) have opposite signs and so the Hall effect acts anti-diffusively. If the Hall effect can overcome diffusive processes, any increment in the local magnetic flux continues to grow and contract until $\langle B_z \rangle$ becomes large enough for $M \rightarrow 0$. By flux conservation, there must be accompanying patches of decreased magnetic flux, which we anticipate having low levels of turbulent transport as well. We associate this scenario with the transition to the LTS.
 
\begin{figure}
\centering
\includegraphics[width=8cm,clip]{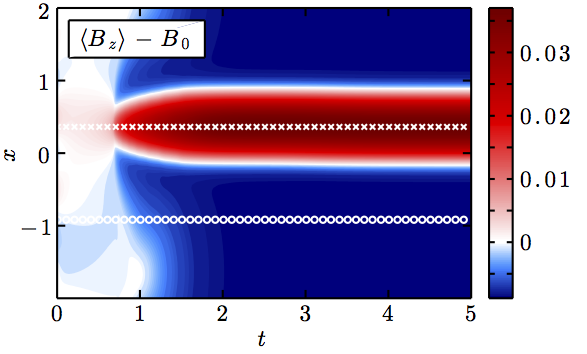}
\newline\newline
\includegraphics[width=8cm,clip]{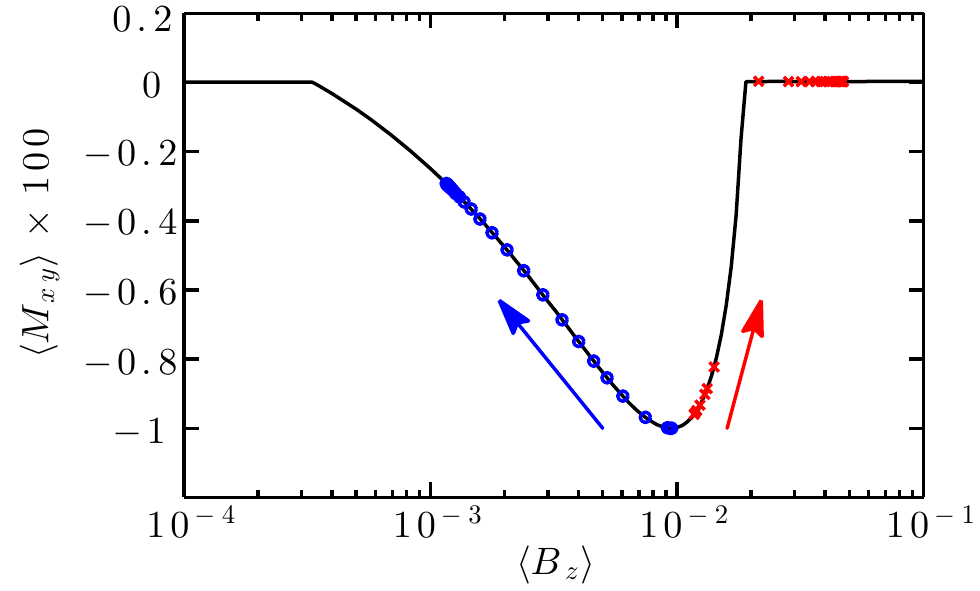}
\newline
\caption{({\it top}) Space-time diagram of the vertically and azimuthally averaged vertical component of the magnetic field $\langle B_z\rangle$ in a toy model (\S\,\ref{sec:case1}) based upon our mean-field theory. The appearance of a zonal magnetic field is observed. ({\it bottom}) Regions of increased field strength (red crosses) evolve anti-diffusively towards $\langle M_{xy}\rangle$ = 0, whereas regions of decreased field strength (blue circles) evolve diffusively towards a low-transport state.}
\label{fig:toy}
\end{figure}

We now make these ideas concrete by specifying a simple model for the Maxwell stress,
\begin{equation}\label{eqn:maxmodel}
M = -0.01 (\gamma / \gamma_{\rm max} ) ,
\end{equation}
where $\gamma$ is the growth rate obtained by solving the dispersion relation (eqs \ref{eqn:growthchannel}--\ref{eqn:phitheta}) with $K = 2 \pi H^{-1} $, $B_0 = \langle B_z \rangle$, and $A = -3/4$; its dependence on $\langle B_z \rangle$ is shown in the bottom panel of Figure \ref{fig:toy} as the solid line. While we do not advocate such a crude relationship between the nonlinear Maxwell stress and the linear properties of the Hall-MRI, this model does satisfy all of the qualitative expectations for $M$ highlighted in the previous two paragraphs. (One could have equally well approximated $M$ by a non-positive function quadratic in $\langle B_z \rangle$ with upwards concavity.)

Using equation (\ref{eqn:maxmodel}), we solve equation (\ref{eqn:bzavg2}) in a periodic domain of length $L_x = 4$, in which a uniform vertical magnetic field of strength $\langle B_z \rangle = 0.01$ is initially disturbed by low-amplitude white noise.\footnote{In order to regularise the solutions of our mean-field equations, we have used a hyper-resistivity $\propto$$k^4$ to damp small-scale growing modes. This extra term mimics the role played by the $z$ direction, which would destabilise and damp structures with radial lengths $\lesssim$$H$.} The result is graphically presented in the top panel of Figure \ref{fig:toy} and bears a striking resemblance to the results of the fully nonlinear numerical simulations (top panel of Fig.~\ref{fig:st-zb1h1}). A zonal magnetic-field is produced, with accompanying regions of low magnetic-field strength. 

To quantify the evolution of $\langle B_z \rangle$, data are collected at fixed intervals in time along lines of constant $x$ where the magnetic field achieves its global maximum and minimum; these points are marked in the top panel of Figure \ref{fig:toy} by the white crosses and white circles, respectively. These data are mapped onto the solid line (eq.~\ref{eqn:maxmodel}) in the bottom panel of Figure \ref{fig:toy}. Regions of locally increased magnetic-field strength (red crosses) find themselves associated with anti-diffusive transport (i.e.~${\rm d}M / {\rm d}\langle B_z \rangle > 0$), further increasing their magnetic-field strength and moving to the right along the curve until  $M = 0$. By contrast, regions of locally decreased magnetic-field strength (blue circles) find themselves associated with diffusive transport (i.e.~${\rm d}M / {\rm d}\langle B_z \rangle < 0$), further decreasing their magnetic-field strength and moving to the left along the curve as $M \rightarrow 0$. Note that $M \ne 0$ in the low-field region; otherwise, the necessary inflection point in $M(x)$ would vanish and the zonal field would diffuse away.

%
%

%
%
\subsection{An analytical criterion for the low-transport state}

From these results, we deduce that a transition from a HTS to a LTS can occur provided that
\begin{equation}\label{eqn:lts}
\frac{c}{e n_{\rm e}} \left. \deriv{\langle B_z \rangle}{M} \right|_{\langle B^0_z \rangle} > \eta + \eta_{\rm t}
\end{equation}
for some $\langle B^0_z \rangle$. To make this criterion more quantitative, we must estimate the size of each of these terms in the HTS. First, we neglect molecular resistivity, which we assume to be small compared to the turbulent resistivity. Second, we assume that the turbulent resistivity is related to the turbulent transport via $\eta_{\rm t} \sim \overline{\alpha} \,\Omega H^2 / {\rm Pm}_{\rm t}$, where ${\rm Pm}_{\rm t} \sim 2$ is the turbulent Prandtl number estimated for this particular component of the turbulent resistivity by \cite{ll09}.\footnote{See also \citet{gg09} and \citet{fs09} for other magnetic-field configurations.} Finally, we approximate ${\rm d} M / {\rm d} \langle B_z \rangle$ by $\overline{\alpha}  \,\rho ( \Omega H)^2 / B_{z,{\rm stab}}$, using the definition of $\overline{\alpha}$ and introducing $B_{z,{\rm stab}}$ as the critical magnetic-field strength above which the longest-wavelength MRI mode is stabilised by magnetic tension. Using these estimates, the bifurcation condition (\ref{eqn:lts}) reduces to
\begin{equation}
\lh \gtrsim \frac{v_{\rm A,stab}}{\Omega \,{\rm Pm}_{\rm t}} ,
\end{equation}
where $v_{\rm A,stab} \equiv B_{z,{\rm stab}} / ( 4 \pi \rho )^{1/2}$. Solving the linear dispersion relation for $v_{\rm A,stab}$, our criterion becomes $\lh \gtrsim 0.2H$, tantalisingly close to the value deduced from Figure \ref{fig:alpha-lambdaH}.

%
%
\subsection{Case II: $\langle M_{xy} \rangle = M( \langle B_z \rangle , \langle \omega_z \rangle)$}\label{sec:case2}

\begin{figure}
\centering
\includegraphics[width=8cm,clip]{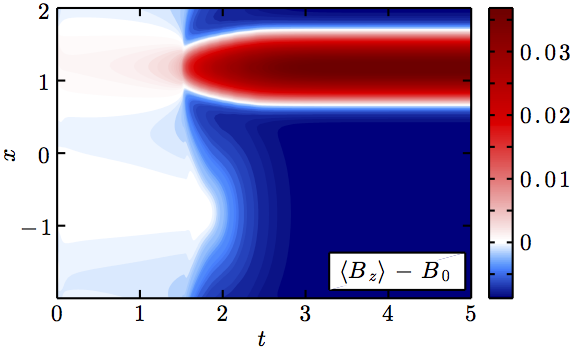}
\newline
\includegraphics[width=8cm,clip]{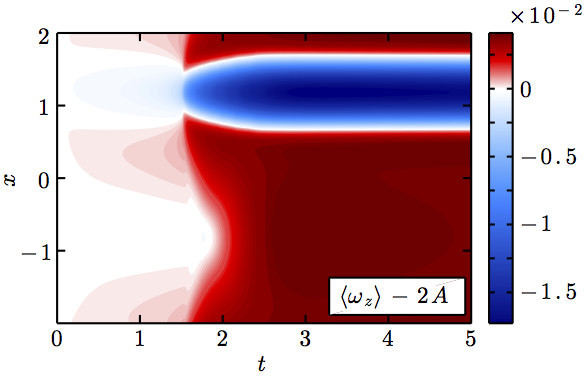}
\newline
\caption{Space-time diagrams of the vertically and azimuthally averaged vertical component of the magnetic field $\langle B_z \rangle$ ({\it top}) and the flow vorticity $\langle \omega_z \rangle$ ({\it bottom}) in a toy model (\S\,\ref{sec:case2}) based upon our mean-field theory. A zonal-vorticity band emerges anti-correlated with the zonal magnetic field, consistent with behaviour found in our numerical simulations (see Fig.~\ref{fig:st-zb1h1}).}
\label{fig:toy2}
\end{figure}

Equation (\ref{eqn:wzavg2}) indicates that the gradient of the Maxwell stress also affects the mean vorticity. While the model for the Maxwell stress presented in Section \ref{sec:case1} is successful at explaining the bifurcation from an HTS to an LTS, it does not take into account this effect, nor does it take into account the feedback of a vorticity-dependent Maxwell stress on the evolution on the magnetic field. Here, we generalise the form of the Maxwell stress to allow for this interplay:
\[
\langle M_{xy} \rangle = M ( \langle B_z \rangle , \langle \omega_z \rangle ) .
\]
To leading order in the perturbation amplitudes, equations (\ref{eqn:bzavg2}) and (\ref{eqn:wzavg2}) then become
\begin{eqnarray}\label{eqn:dbzavg2}
\lefteqn{
\D{t}{\langle B_z^1 \rangle} \simeq  \left( \eta + \eta_{\rm t} - \frac{c}{e n_{\rm e}} \D{\langle B_z \rangle}{M} \right) \DD{x}{\langle B_z^1 \rangle} - \frac{c}{e n_{\rm e}} \D{\langle \omega_z \rangle}{M} \DD{x}{\langle \omega^1_z \rangle} ,
}\nonumber\\*&&\mbox{}
\end{eqnarray}
\begin{eqnarray}\label{eqn:dwzavg}
\lefteqn{
\D{t}{\langle \omega_z^1 \rangle} \simeq  \left( \nu + \nu_{\rm t} + \frac{1}{\rho} \D{\langle \omega_z \rangle}{M} \right) \DD{x}{\langle \omega_z^1 \rangle} + \frac{1}{\rho} \D{\langle B_z \rangle}{M} \DD{x}{\langle B^1_z \rangle} ,
}\nonumber\\&&\mbox{}
\end{eqnarray}
where the partial derivatives of $M$ are evaluated at $\langle B^0_z \rangle$ and $\langle \omega^0_z \rangle$. In principle, the Lorentz and Hall terms may be diffusive or anti-diffusive, depending upon the local gradient of the Maxwell stress. In practise, $\partial M / \partial \langle \omega_z \rangle$ is generally non-negative for MRI-driven turbulence \citep[e.g.][]{pcp08}, and so the Lorentz force acts diffusively on the mean vorticity. As in the previous Section, $\partial M / \partial \langle B_z \rangle$ is positive above some critical field strength and negative below it. We capture this physics by taking the Maxwell stress to be proportional to the channel growth rate, allowing for its dependence on vorticity by promoting $2A \rightarrow \langle \omega_z \rangle$ in the dispersion relation. We then solve equations (\ref{eqn:bzavg2}) and (\ref{eqn:wzavg2}) as before, with $\langle \omega^0_z \rangle = 2A$. The result is graphically presented in Figure \ref{fig:toy2}. A zonal magnetic-field is produced, with the flow vorticity being anti-correlated with the vertical magnetic field. This model captures all of the salient features of the LTS.

%
%

%
%
\section{Discussion}\label{sec:discussion}

In this Paper, we have described the saturation of the MRI in a plasma dominated by the Hall effect. This situation is of particular importance to protoplanetary discs, which are poorly ionised and in which one expects a particularly strong Hall effect near the disc midplane \citep[e.g.][]{wardle07}. A strong Hall effect is considered to be particularly important to the disc dynamics, as it can destabilise plasmas which would otherwise be stabilised by Ohmic losses and could therefore eliminate (or at least significantly reduce) purported `dead zones' in such discs \citep{ws12}.

Using the 3D spectral code Snoopy, we recovered the linear properties of the Hall-dominated MRI. We also demonstrated analytically that MRI `channel' modes remain nonlinearly exact solutions notwithstanding the complicating influence of the Hall effect, and we explored their stability with respect to secondary parasitic modes. We subsequently confirmed the numerical results of SS02, which correspond to the weak-Hall regime ($\lh \sim 10^{-2} H$). By extending those authors' numerical experiments into the Hall-dominated regime ($\lh \gtrsim 0.2 H$), we arrived at our most interesting and perhaps most important finding: the presence of a completely new saturation mechanism at work in MRI-driven turbulence. We have shown that this new saturation mechanism is associated with a reduction in turbulent transport by at least two orders of magnitude compared to `classical' magnetorotational turbulence.

This low-transport state is produced by a redistribution of the vertical magnetic flux into long-lived, narrow ($\sim$$H$), axisymmetric regions, outside of which the mean field averages almost to zero. We refer to these regions as zonal magnetic fields. Coincident with these zonal-field regions are zonal flows, which we have shown to result from (approximate) conservation of canonical vorticity. The accumulation of vertical magnetic flux into zonal-field regions increases the local magnetic tension enough to stabilise their internal flow; outside of these regions, very weak magnetorotational turbulence persists. In fact, the surrounding weak turbulence actively reinforces the integrity of the zonal-field structure by contributing a confining azimuthal electric field. This reinforcement is necessary in the face of resistive (both molecular and turbulent) diffusion.

To explain this behaviour, we have developed a mean-field model based on $y$-$z$ averages of the full set of resistive Hall-MHD equations in the shearing sheet. The key ingredient of this model is the introduction of the $xy$-component of the Maxwell stress $M_{xy}$ into the induction equation by the Hall electric field. This term introduces a new type of radial magnetic diffusion into the averaged induction equation, which can be either diffusive or anti-diffusive depending on the sign of ${\rm d}\langle M_{xy}\rangle /{\rm d}\langle B_z\rangle$. Since both very weak and very strong magnetic fields quench turbulent transport by the MRI (the former through Ohmic losses, the latter by strong tension), we conclude that there will always be regions of MRI-driven turbulence where ${\rm d}\langle M_{xy} \rangle / {\rm d} \langle B_z \rangle > 0$ and therefore the Hall effect will act anti-diffusively. When $\lh \gtrsim 0.2H$, this anti-diffusive behaviour appears to overtake resistive and turbulent diffusion. We have constructed a toy model based upon our mean-field theory that reproduces the zonal-flow structure revealed by our simulations and captures all the salient features of the low-transport state.

Our results indicate that, despite being strongly linearly unstable, Hall-dominated accretion discs could undergo a global bifurcation that substantially reduces the amount of MRI-driven turbulent transport. This contradicts the recent suggestion by \citet{ws12} that the Hall effect could revive turbulent `dead' zones long suspected to exist in the poorly ionised interiors of protoplanetary discs \citep[e.g.][]{gammie96}. Our results also bring into question numerous attempts to estimate the size of the dead zone by coupling chemical networks with MRI stability criteria, whether they be obtained through linear analyses or by naively extrapolating results from resistive-MHD simulations into the Hall-dominated regime. Since strong Hall diffusion can nonlinearly render magnetically active regions magnetorotationally `dead', existing estimates for the extent of the active and dead zones may require revision.

Despite the simplicity of this result and the very general physical principles upon which it is based (e.g.~field-strength-dependent turbulent transport, conservation of canonical vorticity), it is prudent to remind the reader that these results have been obtained using a crude representation (an incompressible and unstratified shearing box) of an actual protoplanetary disc. Since the magnetic quenching responsible for the stabilisation of zonal fields is dependent upon the vertical scale-height, which is artificially set in our simulations to be the vertical size of the shearing box, vertical stratification could affect the dynamics, extent, and magnetic-field strength of the zonal structures. That being said, stratified discs do exhibit a natural length-scale that stabilises MRI modes \citep{gb94,lfg10,lfo13} so that, in principle, zonal fields and flows should also be produced and sustained in stratified shearing boxes. Compressibility, ambipolar diffusion, and the presence of dust grains introduce additional complications by altering the ionisation fraction and the turbulent response of the disc in height-dependent ways.\footnote{Preliminary simulations in unstratified boxes indicate that compressibility and strong ambipolar diffusion do not qualitatively change our results.} Simulations of layered discs (without the Hall effect or ambipolar diffusion) have even shown that magnetorotational turbulence in active surface layers could drive a Reynolds stress \citep{fs03,oml09} and/or a large-scale Maxwell stress \citep{ts08} in the dead zone. We intend to explore the impact of such complexities on the robustness of our results in subsequent work.

We close by remarking that the presence of zonal magnetic fields and zonal flows is of great interest in its own right. While zonal flows have been previously found to naturally emerge in both ideal and resistive simulations of MRI-driven turbulence, those structures are generally weak in amplitude and are believed to be generated by random contributions of the Maxwell stress \citep{jyk09}. The zonal structures we find in Hall-dominated magnetorotational turbulence, on the other hand, are strong in amplitude and are driven by a coherent Maxwell stress acting in concert with conservation of canonical vorticity. A natural suspicion is that these dominant zonal structures may act as particle-trapping sites, enabling fast planetesimal formation through a gravitationally unstable dust layer. The enticing link between Hall-induced turbulent bifurcation to a low-transport state, the occurrence of zonal fields and flows, and the formation of planetesimals will be the subject of a forthcoming publication (Lesur \& Kunz, in preparation).

\section*{Acknowledgments}

Support for M.W.K. was provided by NASA through Einstein Postdoctoral Fellowship Award Number PF1-120084, issued by the Chandra X-ray Observatory Center, which is operated by the Smithsonian Astrophysical Observatory for and on behalf of NASA under contract NAS8-03060. G.L. acknowledges support by the European Community via contract PCIG09-GA-2011-294110. This work was granted access to the HPC resources of IDRIS under allocation x2013042231 made by GENCI (Grand Equipement National de Calcul Intensif). Some of the computations presented in this Paper were performed using the CIMENT infrastructure (https://ciment.ujf-grenoble.fr), which is supported by the Rhone-Alpes region (GRANT CPER07\_13 CIRA: http://www.ci-ra.org). The Texas Advanced Computer Center at The University of Texas at Austin also provided HPC resources under grant number TG-AST090105. This work used the Extreme Science and Engineering Discovery Environment (XSEDE), which is supported by NSF grant OCI-1053575. The authors would like to thank Nuno Loureiro, Jake Simon, and Jim Stone for useful conversations, as well as the expert referee for a prompt and constructive report.

\appendix

\section{Channel stability and parasites}\label{app:parasite}

\subsection{Formulation of the problem and method of solution}

The fact that the Hall-MRI channel flows are exact nonlinear solutions allows one to analytically examine their stability to parasitic modes. One simply considers the channel solution to be part of a time-dependent background state, upon which small-amplitude perturbations are applied. Unfortunately, this time-dependence complicates matters somewhat, since one cannot Fourier decompose in time. We circumvent this difficulty by assuming that the growth rate $\sigma$ of the parasites is much greater than the growth rate $\gamma \sim \Omega$ of the `background' channel solution \citep{gx94}. Since $\sigma \sim b\Omega$, this amounts to the assumption that the channel mode has grown to large amplitude ($b \gg 1$) and, from the perspective of the parasites, may be considered a stationary equilibrium. This ordering implies that the effects of rotation and shear on the parasites, as well as the uniform background magnetic field $\bb{B}_0$, may be ignored. What results is an 8th-order boundary-value problem in $z$ with eigenvalue $\sigma$.

Following this prescription, we substitute
\[
\bb{v} = \bb{v}_{\rm ch} + \delta \bb{v} , \quad \bb{B} = \bb{B}_{\rm ch} + \delta \bb{B} , \quad P = P_0 + \delta P
\]
into equations (\ref{eqn:force})--(\ref{eqn:divv}), linearise in the perturbation amplitudes, and search for Fourier modes with the space-time dependence $\delta \propto \exp( \sigma t + \imag \bb{k} \bcdot \bb{x} )$. The perturbation wavevector
\[
\bb{k} = k  \left( \ex \sin \theta_k - \ey \cos \theta_k \right)
\]
is parallel to the channel magnetic (velocity) field when $\theta_k = \theta$ ($ \theta_k = \phi \pm \pi / 2$). The resulting set of equations is
\begin{eqnarray}\label{eqn:parforce}
\lefteqn{
\sigma \delta \bb{v} = -\imag \bb{k} \bcdot \bb{v}_{\rm ch} \, \delta \bb{v} - \delta v_z \deriv{z}{\bb{v}_{\rm ch}} - \frac{1}{\rho} \left( \imag \bb{k} + \ez \deriv{z}{} \right) \delta \Pi
}\nonumber\\*&&\mbox{}
+ \imag \bb{k} \bcdot \bb{B}_{\rm ch} \, \frac{ \delta \bb{B}}{4\pi\rho} + \frac{\delta B_z}{4\pi\rho} \deriv{z}{\bb{B}_{\rm ch}} + \nu \left( \dderiv{z}{} - k^2 \right) \delta \bb{v} ,\end{eqnarray}
\begin{eqnarray}\label{eqn:parinduction}
\lefteqn{
\sigma \delta \bb{B} = -\imag \bb{k} \bcdot \left(  \bb{v}_{\rm ch} - \frac{\bb{J}_{\rm ch}}{e n_{\rm e}} \right) \delta \bb{B} - \left( \delta v_z - \frac{\delta J_z}{e n_{\rm e}} \right) \deriv{z}{\bb{B}_{\rm ch}} 
}\nonumber\\*&&\mbox{}
+ \imag \bb{k} \bcdot \bb{B}_{\rm ch} \left( \delta \bb{v} - \frac{\delta \bb{J}}{en_{\rm e}} \right) + \delta B_z \deriv{z}{} \left( \bb{v}_{\rm ch} - \frac{\bb{J}_{\rm ch}}{en_{\rm e}} \right)
\nonumber\\*&&\mbox{}
+ \eta \left( \dderiv{z}{} - k^2 \right) \delta \bb{B} ,
\end{eqnarray}
\begin{equation}\label{eqn:pardivv}
\imag \bb{k} \bcdot \delta \bb{v} + \deriv{z}{\delta v_z} = 0 ,
\end{equation}
where
\[
\delta \Pi = \delta P + \frac{ \bb{B}_{\rm ch} \bcdot \delta \bb{B} }{ 4\pi }
\]
is the perturbation of the total pressure and $\bb{J}_{\rm ch}$ is the channel current density (eq.~\ref{eqn:jch}).

The linear operator associated with equations (\ref{eqn:parforce})--(\ref{eqn:pardivv}) is periodic in $L \equiv 2 \pi K^{-1}$, and we thus make the Floquet ansatz $\delta \propto \delta(z) \exp( \imag k_z z )$ where the function $\delta(z)$ is $L$-periodic. Following \citet{llb09}, we solve the resulting set of equations numerically via a pseudospectral technique: the $z$-domain is partitioned into $N$ ($=$$256$) grid points and the operator is discretised using Fourier cardinal functions \citep[see][]{boyd00}. This procedure leads to a $7N \times 7N$ generalised algebraic eigenvalue problem, which we solve using the QZ algorithm. The resulting (complex) eigenvalues are the growth rates $\sigma$ of the eigenmodes.

In the following Sections, we give representative examples of three classes of parasitic eigenmodes. First, we take $\bb{k} \!\perp\! \bb{B}_{\rm ch}$ and $\eta = 0$. The resulting `kink' and `kink-pinch' modes are essentially hydrodynamical disturbances of the channel jets, whose growth is afforded by the shear free energy stored in the channel mode. Second, we take $\bb{k}\, || \,\bb{B}_{\rm ch}$ and $\eta \ne 0$. The resulting `pinch-tearing' mode extracts little to no free energy from the flow shear, and instead relies on Ohmic dissipation to facilitate the reconnection of neighbouring field lines of opposite polarity and thereby tap into the magnetic energy stored in the channel mode. Finally, we take $\bb{k} \, || \, \bb{B}_{\rm ch}$ and $\eta = 0$ and investigate whether the Hall-shear instability can act as a parasitic mode by feeding off the shear free energy of the channel. In all cases we set $\nu = 0$ and $\Omega / \omega_{\rm H,0} = 17.4$, the latter of which corresponds to our fiducial simulation ZB1H1.

\subsection{Kink and kink-pinch parasites}

The kink (i.e.~Type I) and kink-pinch (i.e.~Type II) parasitic modes are described in detail in \citet[][\S\,3.4]{gx94} and \citet[][\S\,2.3.3]{llb09}, to which we refer the reader. These parasites are most clearly identified when $\eta = 0$ and $\bb{k} \! \perp \! \bb{B}_{\rm ch}$, conditions which simultaneously alleviate the stabilising influence of magnetic tension and maximise the access to the shear free energy of the channel. In brief, the kink mode is a Kelvin-Helmholtz instability feeding upon the inflection points in the channel velocity profile. It is associated with $k_z = 0$ and, as such, is the dominant parasite for a two-stream channel. The kink-pinch mode is a hybrid mode, exhibiting both kink-like and pinch-like characteristics. Rather than exhibiting a phase velocity that matches the inflection points of the channel velocity profile, the kink-pinch mode appears to influence the flow most strongly near the points in $z$ where the channel magnetic field changes sign. This mode grows about an order of magnitude slower than the pure kink mode and, having $k_z \ne 0$, only afflicts channels with more than two streams.

\begin{figure}
\centering
\includegraphics[width=8cm,clip]{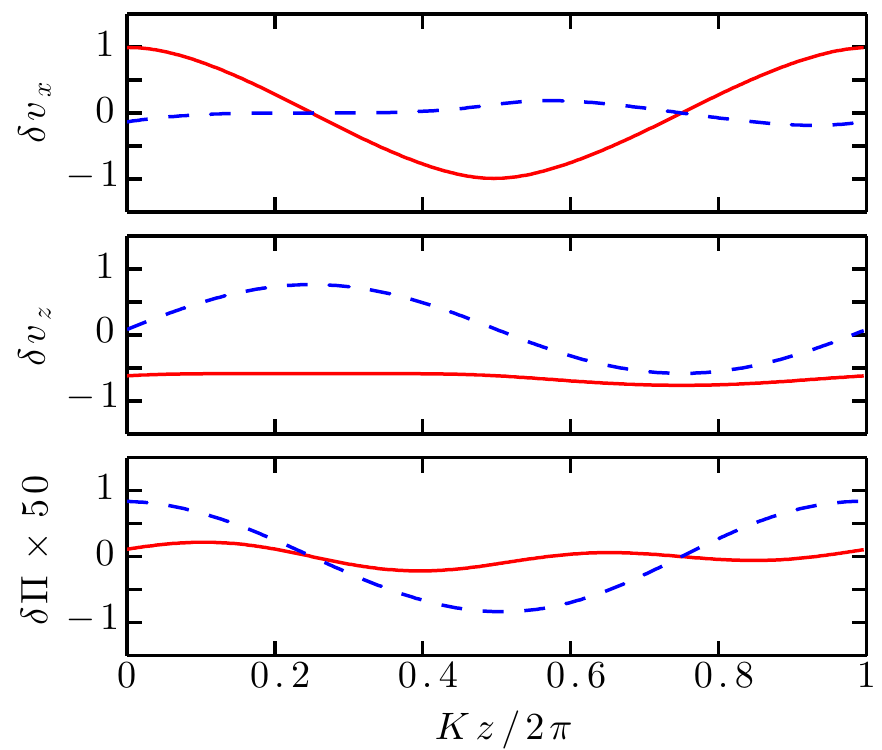}
\newline\newline
\includegraphics[width=8cm,clip]{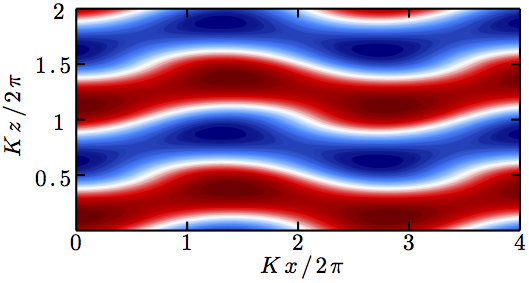}
\newline
\caption{({\em top}) The $\delta v_x$, $\delta v_z$, and $\delta \Pi$ components of the eigenfunction of a kink mode with $\theta = \phi = \pi / 4$, $k / K = 0.5$, $\theta_k = -\pi / 4$, and $k_z = 0$. The solid (dashed) lines denote the real (imaginary) parts. The total eigenfunction is normalised so that ${\rm max} | \delta v_x | = 1$. The growth rate $\sigma / b \Omega = 0.008616$. ({\em bottom}) Coloured iso-contours of the real part of $v_x$ at $y=0$ in the $(x,z)$ plane. The background is a four-stream Hall-MRI channel with jets centred at $Kz = n\pi /2$ with $n = 1,~3,~5$ and $7$. The perturbation is normalised so that ${\rm max} | \delta v_x | = v_{\rm ch}$. }
\label{fig:kink}
\end{figure}

In the top panel of Figure \ref{fig:kink} we plot the $\delta v_x$, $\delta v_z$, and $\delta \Pi$ components of the eigenfunction of a kink mode with $\theta = \phi = \pi / 4$, $k / K = 0.5$, $\theta_k = -\pi/4$, and $k_z = 0$. The solid (dashed) lines denote the real (imaginary) parts. In the bottom panel, we display coloured iso-contours of the real part of $v_x$ at $y=0$ in the $(x,z)$ plane. The Hall effect does not significantly impact the structure of the mode since $\bb{k} \bcdot \bb{B}_{\rm ch} = 0$. Just as in ideal MHD, a pressure gradient develops across each jet and gives rise to a kink motion, deflecting the channels upwards or downwards. The Hall effect does, however, reduce the parasite growth rate from $\sigma / b \Omega = 0.1959$ to $0.008616$. This is because a strong Hall effect with $\Omega / \omega_{\rm H,0} > 0$ substantially decreases the shear free energy $K v_0$ stored in the channel mode (see eq.~\ref{eqn:kv0channel}).

%
%

\begin{figure}
\centering
\includegraphics[width=8cm,clip]{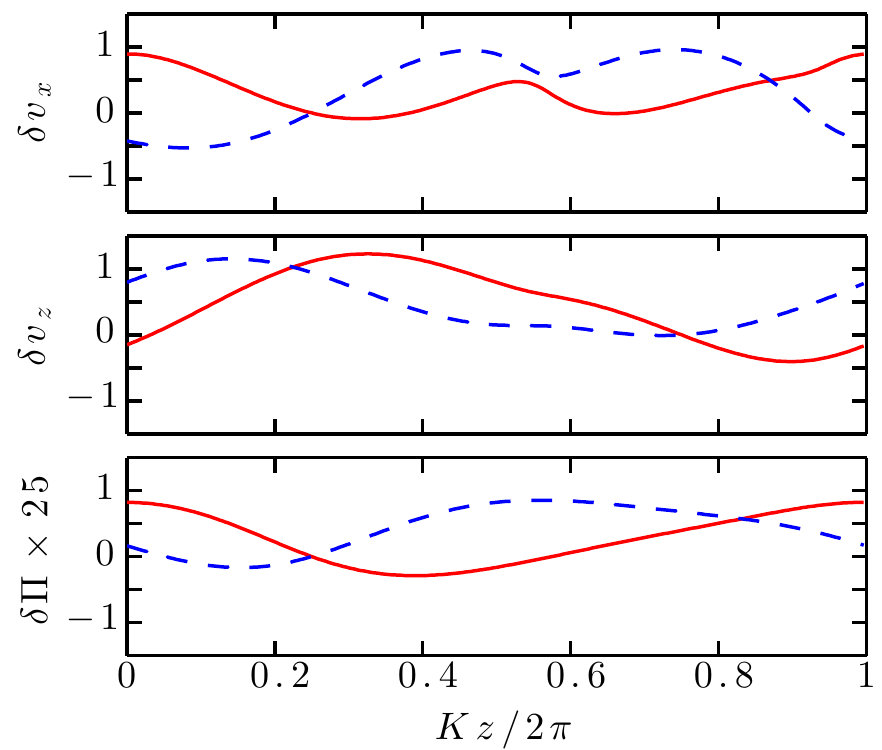}
\newline\newline
\includegraphics[width=8cm,clip]{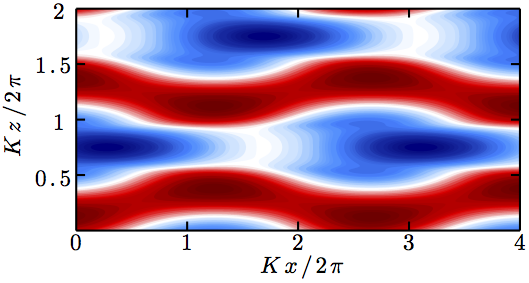}
\newline
\caption{({\em top}) The $\delta v_x$, $\delta v_z$, and $\delta \Pi$ components of the eigenfunction of a kink-pinch mode with $\theta = \phi = \pi / 4$, $k / K = 0.5$, $\theta_k = -\pi / 4$, and $k_z = 0.5$. The solid (dashed) lines denote the real (imaginary) parts. The total eigenfunction is normalised so that ${\rm max} | \delta v_x | = 1$. The growth rate $\sigma / b \Omega = 0.004093 + 0.0058 \imag$. ({\em bottom}) Coloured iso-contours of the real part of $v_x$ at $y=0$ in the $(x,z)$ plane. The background channel and normalisation are as in Figure \ref{fig:kink}. The entire pattern is moving to the left because $\sigma$ possesses a positive imaginary part.}
\label{fig:kinkpinch}
\end{figure}

We find a similar situation for the kink-pinch mode ($k_z = 0.5$), shown in Figure \ref{fig:kinkpinch}. As in ideal MHD, the pressure perturbation $\delta \Pi$ changes sign at $Kz = \pi/2$ and kinks the jet centred there, while the vertical velocity perturbation $\delta v_z$ changes sign at $Kz = 3 \pi / 2$ and pinches the jet centred there. This gives rise to the alternate kinking and pinching of neighbouring jets, which can be observed in the bottom panel of the figure. For the same reason as for the kink mode, the growth rate decreases from $\sigma / b \Omega = 0.0931 + 0.1319 \imag$ to $0.004093 + 0.0058 \imag$. The complex conjugate of this mode has the pressure perturbation kink the $Kz = 3\pi / 2$ jet and the vertical velocity perturbation pinch the $Kz = \pi / 2$ jet.

For both the kink and kink-pinch modes, the growth rates are much too small to be of relevance for the saturation of the Hall-dominated MRI. In order for $\sigma \sim \gamma$, the usual criterion for determining when secondary parasitic modes might overtake the primary channel mode and instigate a break-down into 3D turbulence \citep{pg09}, the channel must grow to amplitudes $b \sim 100$. By this stage, compressibility effects become important, a feature not captured in our parasite linear analysis nor in our nonlinear simulations.

%
%

\subsection{Pinch-tearing parasite}

\begin{figure}
\centering
\includegraphics[width=8cm,clip]{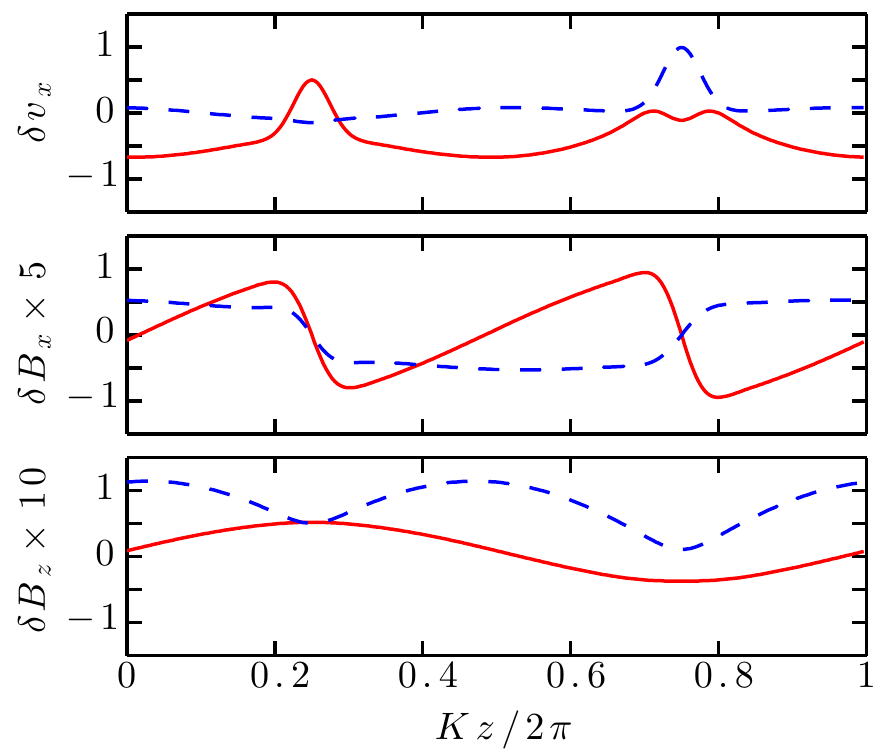}
\newline\newline
\includegraphics[width=8cm,clip]{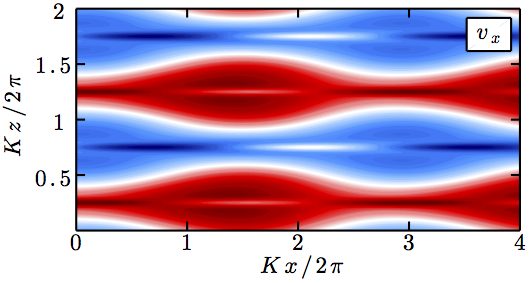}
\newline\newline
\includegraphics[width=8cm,clip]{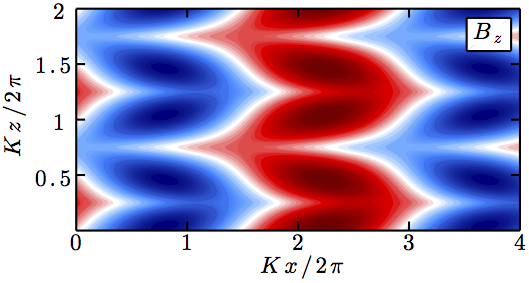}
\newline
\caption{({\em top}) The $\delta v_x$, $\delta B_x$, and $\delta B_z$ components of the eigenfunction of a pinch-tearing mode with $\theta_k = \pi / 3$, $k / K = 0.4$, $k_z = 0$, and $b=5$. The channel orientation $\theta \simeq 0.255\pi$ is obtained by solving the dispersion relation for the fastest-growing mode with $\beta$, $\Lambda_{\rm H}$, and $\Lambda_\eta$ taken from our fiducial simulation ZB1H1. The solid (dashed) line denotes the real (imaginary) part. The total eigenfunction is normalised so that ${\rm max} | \delta v_x | = 1$. The growth rate $\sigma / b \Omega = 0.02868 - 0.03449 \imag$. ({\em bottom}) Coloured iso-contours of the real parts of $v_x$ and $B_z$ at $y=0$ in the $(x,z)$ plane. The background channel and normalisation are as in Figure \ref{fig:kink}. The entire pattern is moving to the right because $\sigma$ possesses a negative imaginary part.}
\label{fig:pinch}
\end{figure}

When resistivity is included, pinching motions are subject to tearing instabilities driven by magnetic reconnection. This not only modifies the kink-pinch mode by opening a pathway to extract the extra energy stored in the channel's magnetic field, but also facilitates the growth of pinch-tearing modes (see \S\,2.3.4 of \citealt{llb09}). The latter manifest most readily when $\theta_k = \theta$, i.e.~when the wavevector and the channel magnetic field are aligned. As $\theta_k$ deviates from $\theta$, the pinch-tearing mode bifurcates into two complex-conjugate modes, each one localised on one or the other jet. This is the situation we display in Figure \ref{fig:pinch}, which highlights the $\delta v_x$, $\delta B_x$, and $\delta B_z$ components of the eigenfunction of a pinch-tearing mode with $\theta_k = \pi / 3$, $k / K = 0.4$, $k_z = 0$, and $b = 5$. The eigenmode localises preferentially on the lower jet, with a growth rate $\sigma / b \Omega = 0.02868 - 0.03449 \imag$; its complex conjugate localises on the upper jet. 

While the Hall effect with $\Omega / \omega_{\rm H,0} > 0$ decreases the magnetic shear energy stored in the channel, it also generally increases the growth rate of the pinch-tearing mode. Indeed, this parasite grows faster than both the kink and kink-pinch modes described above by a factor $\sim$$10$. Nevertheless, its growth rate remains quite small; in simulations of the Hall-dominated MRI in which a two-stream $k_x = 0$ channel mode is deliberately excited, this parasite grows over such a long timescale that the shearing of its wavevector (a feature not taken into consideration here) wraps the mode into an almost axisymmetric configuration with $k > 1$.\footnote{We refer the reader to Appendix B of \citet{lfg10} for a quantitative assessment of how shear retards parasite growth.} In fact, the channel grows to such large amplitudes ($b > 300$) that the numerical timestep drops precipitously and further time integration becomes untenable. In simulations where a $k_x = 0$ channel mode is not exclusively excited by hand (e.g.~all of those listed in Table \ref{tab:runs}), the linear phase is disrupted at $b \sim 10$, not by parasites, but rather by the interference of large-amplitude $k_x \ne 0$ channels.

%
%

\subsection{Hall-shear instability as a channel parasite}

In Section \ref{sec:test} we employed the Hall-shear instability (HSI) as a test of our numerical algorithm. In the context of the channel mode, the background shear is provided by the sinusoidal variation of $v_{\rm ch}$ and is strongest at the vertical locations satisfying $Kz = n\pi$ with $n = 0$, $\pm 1$, $\pm 2$, $\dots$. The magnitude of the channel magnetic field is also greatest at those heights. As a result, any local patch centred on $Kz = n\pi$ resembles a non-rotating shearing sheet with a vorticity-aligned magnetic field---precisely the setup considered by K08\nocite{kunz08}. While this analogy suggests that the HSI could act as a powerful channel parasite, this expectation is not borne out in our numerical simulations. Here we adapt the K08 calculation to determine why this is so.

The HSI is most easily examined in the limit of negligible resistivity, in which case the channel magnetic and velocity fields are mutually perpendicular ($\theta = \phi$; see eq.~\ref{eqn:phitheta}). We can therefore erect an orthonormal coordinate system oriented with the channel mode: $\eb = \bb{B}_{\rm ch} / B_{\rm ch}$, $\ev = \bb{v}_{\rm ch} / v_{\rm ch}$, and $\ez = \eb \btimes \ev$. In this geometry, wavevectors parallel to the channel magnetic field ($\bb{k} = k \eb$) have the greatest potential for growth. The $z$- and $v$-components of the linearised induction equation (\ref{eqn:parinduction}) become\footnote{These equations may be profitably compared with eqns (46a,b) of K08 \nocite{kunz08} after making the replacements $z \rightarrow x$ and $v \rightarrow y$.}
\begin{equation}\label{eqn:hsiz}
\sigma \delta B_z - b \cos Kz \, \frac{ c k^2 B_0}{4 \pi e n_{\rm e}} \, \delta B_v = \imag k B_0 \,b \cos Kz ~\delta v_z ,
\end{equation}
\begin{eqnarray}\label{eqn:hsiv}
\lefteqn{
\sigma \delta B_v + b \cos Kz \left[ \frac{c k^2 B_0}{4\pi e n_{\rm e}} \left( 1 - \frac{1}{k^2}\dderiv{z}{} - \frac{K^2}{k^2}\right) - K v_0 \right] \delta B_z
}\nonumber\\*&&\mbox{}
= \imag k B_0 \,b \cos Kz ~\delta v_v .
\end{eqnarray}
It is clear from equation (\ref{eqn:hsiv}) that the shear of the channel mode (represented by the final term in the brackets) uses $\delta B_z$ to generate $\delta B_v$. The Hall terms, on the other hand, generate $\delta B_z$ at the expense of $\delta B_v$. This effect is present even in the absence of shear and arises because the $v$-component of the perturbed electron velocity differs from the ion-neutral velocity by
\[
- \frac{\delta J_v}{en_{\rm e}} = \frac{\imag ck}{4\pi e n_{\rm e}} \left( 1 - \frac{1}{k^2} \dderiv{z}{} \right) \delta B_z .
\]
The induced magnetic field is sheared further, and there is the potential for runaway.

It is a straightforward exercise to show from equations (\ref{eqn:parforce}), (\ref{eqn:hsiz}), and (\ref{eqn:hsiv}) that, whether $k \gg K,{\rm d}/{\rm d}z$ (the limit captured by the K08\nocite{kunz08} analysis) or ${\rm d} / {\rm d}z \gg k,K$ (a WKBJ treatment), a necessary condition for instability is
\begin{equation}\label{eqn:hsi}
1 < \frac{ K v_0 } { \omega_{\rm H,0} }.
\end{equation}
Physically, this inequality states that the time required from an ion to execute one orbital gyration around a magnetic-field line must be longer (by a factor of $n_{\rm e} / n$) than the time it takes for a magnetic perturbation to grow by shear. If this condition is not met, the ions are well-coupled to the electrons (and thereby to the magnetic field), and we are left with simple linear-in-time growth due to shearing of the magnetic-field perturbation by the channel flow. (This criterion is analogous to eq.~43 of K08\nocite{kunz08}.)

What complicates matters beyond those investigated in K08 is that here the channel shear, which provides the free energy for growth, is itself a function of $\omega_{\rm H,0}$. Plugging in our expression for the channel shear (eq.~\ref{eqn:kv0channel} with ${\rm Rm_{eff}} \rightarrow \infty$), our instability criterion (\ref{eqn:hsi}) for the HSI parasites becomes equation (\ref{eqn:hmristable}), precisely the stability criterion for the Hall-MRI channels themselves! In other words, if the Hall-MRI channels are active in the disc, then the HSI cannot act as a parasitic instability.

%
%
\section{Numerical stability in Hall-MHD}\label{app:hallstab}

\citet{falle03} suggested that explicit schemes for numerically solving the equations of Hall-MHD are unconditionally unstable due to the existence of small-wavelength whistler waves. Although this conclusion is correct for the numerical schemes \citet{falle03} considered, here we demonstrate that {\em higher-order} time-explicit schemes, such as the one used in Snoopy, are stable without the need for physical (e.g.~Ohmic or ambipolar) or artificial (e.g.~hyper-resistive) wave damping.

We start by considering the induction equation (eq.~\ref{eqn:induction}) with the first (ideal) and third (Ohmic) terms on the right-hand side dropped. Decomposing the magnetic field into a fixed guide field $\bb{B}_0$ and a small-amplitude fluctuation $\delta \bb{B}(t) \exp( \imag \bb{k} \bcdot \bb{x} )$, we find that linear whistler waves are described by
\begin{equation}\label{eqn:lininduction}
\deriv{t}{\delta\bb{B}} = \frac{ c \bb{k} \bcdot \bb{B}_0 }{ 4 \pi e n_e } \; \left( \bb{k} \btimes \delta \bb{B} \right) .
\end{equation} 
In spectral codes such as Snoopy the right-hand side of this equation is computed exactly using Fourier decomposition, and we adopt this scheme in what follows.

Without loss of generality we take the wavevector $\bb{k} = k \ez$ and magnetic-field perturbation $\delta \bb{B} = \delta B_x \ex + \delta B_y \ey$, ensuring $\bb{k} \bcdot \delta \bb{B} = 0$. Equation (\ref{eqn:lininduction}) can then be written as
\begin{equation}\label{eqn:halleqn}
\deriv{t}{\delta\bb{B}} = \msb{R} \, \delta \bb{B} , \quad {\rm where}\quad \msb{R} \equiv \frac{c k^2 B_{0,z} }{ 4 \pi e n_{\rm e} } 
\left( 
\begin{array}{cc}
0 & -1 \\
1 & 0
\end{array}
\right) .
\end{equation}
We integrate equation (\ref{eqn:halleqn}) forward in time from $t^{(n)}$ to $t^{(n+1)}$ using an RK3 scheme similar to that used in Snoopy. For a system of differential equations $\bb{y}' = \bb{f} ( \bb{y} )$, this procedure reads:
\begin{eqnarray}
\bb{q}_1 &=& \bb{f} \left( \bb{y}^{(n)} \right) \nonumber\\*
\bb{q}_2 &=& \bb{f} \left( \bb{y}^{(n)} + \frac{h}{2} \bb{q}_1 \right) \nonumber\\*
\bb{q}_3 &=& \bb{f} \left( \bb{y}^{(n)} - h \bb{q}_1 + 2 h \bb{q}_2 \right) \nonumber\\*
\bb{y}^{(n+1)} &=& \bb{y}^{(n)} + \frac{h}{6} \bigl( \bb{q}_1 + 4 \bb{q}_2 + \bb{q}_3 \bigr) , \nonumber
\end{eqnarray}
where $h \equiv t^{(n+1)} - t^{(n)}$. Applying this algorithm to equation (\ref{eqn:halleqn}), we find
\begin{equation}
\delta \bb{B}^{(n+1)}=\msb{Q} \, \delta \bb{B}^{(n)}
\end{equation}
for
\[
\msb{Q} = \left( 
\begin{array}{cc}
1 - \frac{1}{2} \varepsilon^2       			&  	-\varepsilon + \frac{1}{6} \varepsilon^3  \\
\varepsilon - \frac{1}{6} \varepsilon^3	& 	1 - \frac{1}{2} \varepsilon^2
\end{array} \right) 
\quad {\rm and} \quad \varepsilon \equiv h \, \frac{c k^2 B_{0,z} }{ 4 \pi e n_{\rm e} } .
\]
Note that the matrix $\msb{Q}$ is a third-order expansion of the formal solution $\delta \bb{B}^{(n+1)} = \exp(h \msb{R}) \, \delta \bb{B}^{(n)}$. Extensions to higher order are straightforward.

Stability is guaranteed if the eigenvalues of $\msb{Q}$,
\[
\lambda_\pm = 1 - \frac{\varepsilon^2}{2} \mp \imag \left( \varepsilon - \frac{\varepsilon^3}{6} \right) ,
\]
satisfy the inequality $| \lambda_\pm | < 1$. The numerical scheme is therefore stable provided $\varepsilon < \sqrt{3}$; Snoopy uses $\varepsilon = 1.5$. It can easily be shown by this approach that similar schemes of first or second order in time, such as the ones considered by \citet{falle03}, are unconditionally unstable. The fourth-order Runge-Kutta scheme is stable for $\varepsilon < 2\sqrt{2}$.

In conclusion, the third-order explicit time integrator employed in Snoopy guarantees that linear whistler waves are stable, without the need for additional diffusion terms.

\bibliographystyle{mn2e}

\bibliography{hmri}

\bsp
\label{lastpage}

\end{document}